\documentclass[10pt]{IEEEtran}

\usepackage{color}
\usepackage{echodefs}
\usepackage{todonotes}

\newcommand{\calD}{\mathcal{D}}
\newcommand{\calL}{\mathcal{L}}
\newcommand{\calJ}{\mathcal{J}}
\newcommand{\calN}{\mathcal{N}}
\newcommand{\vectorize}{\mathsf{vec}}

\title{Localizing Unsynchronized Sensors with Unknown Sources}
\author{Dalia El Badawy, Viktor Larsson, Marc Pollefeys, and Ivan Dokmani\'c\thanks{In line with the philosophy of reproducible research, all code and data to reproduce the results of this paper are available at \url{http://github.com/swing-research/xtdoa}.}
\thanks{Work done while D. El Badawy was a student at EPFL, Switzerland.}
\thanks{V. Larsson is with the Department of Computer Science, ETH Zurich.}
\thanks{M. Pollefeys is with the Department of Computer Science, ETH Zurich and Microsoft Mixed Reality and AI Lab Zurich.}
\thanks{I. Dokmani\'c is with the Department of Mathematics and Computer Science, University of Basel.}}
\date{2023}

\begin{document}

\maketitle

\begin{abstract}
We propose a method for sensor array self-localization using a set of sources at unknown locations. The sources produce signals whose times of arrival are registered at the sensors. We look at the general case where neither the emission times of the sources nor the reference time frames of the receivers are known. 
Unlike previous work, our method directly recovers the array geometry, instead of first estimating the timing information. The key component is a new loss function which is insensitive to the unknown timings. We cast the problem as a minimization of a non-convex functional of the Euclidean distance matrix of microphones and sources subject to certain non-convex constraints. After convexification, we obtain a semidefinite relaxation which gives an approximate solution; subsequent refinement on the proposed loss via the Levenberg-Marquardt scheme gives the final locations. 
Our method achieves state-of-the-art performance in terms of reconstruction accuracy, speed, and the ability to work with a small number of sources and receivers. It can also handle missing measurements and exploit prior geometric and temporal knowledge, for example if either the receiver offsets or the emission times are known, or if the array contains compact subarrays with known geometry.
\end{abstract}

\section{Introduction} 
\label{sec:introduction}

In MIMO radar \cite{SanAntonio:2007mimo}, ultrasound imaging \cite{Parhizkar:2013calibration}, underwater acoustics \cite{ferguson1989underwater}, time-reversal \cite{devaney2005, ciuonzo2015}, and room acoustics \cite{scheibler2018separake, pan2017frida} a collection of sources emit signals that are then captured by the receivers. In these applications, we often need an accurate knowledge of the geometry of the source and receiver positions to proceed with common array processing algorithms such as beamforming and source localization. Knowing the sensor array geometry similarly enables the reconstruction of physical fields in environmental monitoring using ad hoc sensor networks \cite{Martinez:2013fukushima,Simoni:2011hydrologic}. Further, knowing device locations enables a host of location-based services in the context of the Internet of Things \cite{win2018efficient}. The recurring quintessential problem is thus to efficiently estimate the geometry of a set of nodes.

\begin{figure*}
\centering
\begin{minipage}[b]{\linewidth}
  \centering
    \centerline{\includegraphics[width=0.75\linewidth]{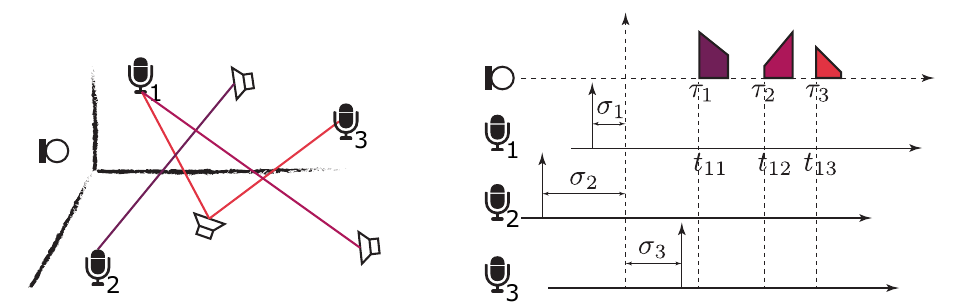}}
\medskip
\end{minipage}%
\caption{The problem setup addressed in this paper. Source events (such as sounds in the environment produced by people or loudspeakers, or electromagnetic signals of opportunity) are captured by the receivers. In this illustration, there are three loudspeakers (sources) with emission times $\{\tau_k\}_{k=1}^{3}$. We also have three microphones (receivers) numbered 1, 2, 3 with internal times of reference $\{\sigma_m\}_{m=1}^{3}$ relative to a global reference (not numbered). The time of arrival $t_{mk}$ at each microphone depends on the source's emission time $\tau_k$ and the microphone's offset $\sigma_m$. Not all source events reach all receivers, thus some entries $t_{mk}$ might be missing. }
\label{fig:setup}
\hfill
\end{figure*}

We study estimating the geometry of the nodes from the times of arrival (TOAs) of the source signals. The setup is illustrated in Figure \ref{fig:setup} where receivers register TOAs of source events. Some events can be (near-)collocated with the nodes, such is the case with transceivers. If the sources are anchors with known positions, locating the nodes becomes an exercise in geometry: intersecting spheres or hyperboloids depending on whether or not the devices are synchronized. However, the most general and practically appealing setup is when the sources are at arbitrary, unknown locations. This enables one to use signals of opportunity such as speech, transient sounds, or radio signals \cite{simkovits2017navigation}. But it also means that receivers are no longer synchronized with the sources. To complicate things further, receivers themselves do not have to be mutually synchronized. While in many audio applications, microphones are connected to a common interface, we are also surrounded by ad hoc networks of smartphones and voice-based assistants. Thus, not only do we measure times of arrival as opposed to absolute times of flight, but those registered times of arrival further depend on the unknown time reference frame of each receiver. In this paper, we introduce an algorithm for jointly localizing a set of receivers and a set of sources from measured TOAs in the general case when all nodes are not synchronized: sources go off at unknown times and the reference time frame of each receiver can be different and unknown.

\subsection{Notation} 
\label{sub:notation}

We denote matrices with capital letters ($\mS$, $\mR$) and vectors with lowercase letters ($\vs$, $\vr$). Uppercase subscripts indicate the size, e.g., $\mJ_{L}$ is an $L\times L$ matrix, and $\mR_{M\times K}$ is an $M\times K$ matrix. Lowercase subscripts $d_{ij}$ indicate the element at the $i^\textrm{th}$ row and $j^\textrm{th}$ column of a matrix. A superscript $T$ as in $\mJ^T$ denotes the matrix transpose. Lowercase superscripts $x^{(k)}$ denote the $k^\textrm{th}$ iteration. The real $d$-space is denoted $\R^d$. The vectorization operator is denoted $\vectorize(x)$. We use the diagonalization operator where $\diag(\mG)$ is a vector of the elements on the diagonal of the matrix. Similarly,  $\diag(\sigma_1,\dots, \sigma_N)$ is an $N\times N$ matrix with $\sigma_1,\dots, \sigma_N$ on the diagonal.

\subsection{Related Work} 
\label{sub:related_work}

Among the earliest work on self-localization are the papers of Rockah and Schultheiss \cite{Rockah:1987array-partI,Rockah:1987array-partII} and Weiss and Friedlander \cite{Weiss:1989array} from the 1980s. Plinge et al. \cite{plinge2016acoustic} provide an in-depth overview of microphone array localization techniques. A  systematization of existing approaches to self-localization is also given by Wang et al. \cite{wang2015self}. In the following, we review some of the major points, related to localization in different setups. An overview is also shown in Table \ref{tab:relatedwork}.

\begin{table*}[!htb]
\caption{Overview of related work with respect to the setup.}
\label{tab:relatedwork}
\centering
\begin{tabular}{llll}
\toprule
Approach & Input & Unknowns & Challenges\\\hline
\multicolumn{1}{p{4cm}}{Multidimensional scaling (MDS) \cite{torgerson1952multidimensional,kruskal1978multidimensional,dokmanic2015euclidean,vanwynsberghe2016jasa}} & Pairwise distances & - & \multicolumn{1}{p{4cm}}{Requires full synchronization} \\
\multicolumn{1}{p{4cm}}{Multidimensional unfolding \cite{Schonemann:1970wd, Crocco:2012eu}, ML optimization \cite{Weiss:1989array}} & Distances between nodes/events & - & \multicolumn{1}{p{4cm}}{Requires full synchronization, Bad local minima}\\
\multicolumn{1}{p{4cm}}{SDP relaxations \cite{biswas2006semidefinite1,biswas2006semidefinite2, yang2009efficient,vaghefi2013asynchronous,wang2012semidefinite,alfakih1999,ding2010}} & Distances/TDOA/FDOA & - & \multicolumn{1}{p{4cm}}{Requires anchor nodes or positions of the sensor nodes} \\
\multicolumn{1}{p{4cm}}{Majorization \cite{ono2009blind}, Two-stage \cite{zhang2016iwaenc,Gaubitch:2013km, thrun2006affine, krekovic2018structure}} & TDOA & Source or receiver offsets & \multicolumn{1}{p{4cm}}{Bad local minima, cannot handle near-minimal configurations}\\
Two-stage \cite{wang2015self} & TDOA & Source \& receiver offsets & \multicolumn{1}{p{4cm}}{Slow, cannot handle near-minimal configurations}\\ 
Proposed &  TOA/TDOA & Source \& receiver offsets  &  - \\

\bottomrule
\end{tabular}
\end{table*}

In some cases, the pairwise distance between all the nodes can be estimated. This happens for example when the nodes can both send and receive \cite{peng2007beepbeep} or by measuring the diffuse noise coherence \cite{vanwynsberghe2016reseaux}. Localization then amounts to multidimensional scaling (MDS) \cite{torgerson1952multidimensional,kruskal1978multidimensional,dokmanic2015euclidean,vanwynsberghe2016jasa}. 

A more common situation in audio applications is that the nodes can only receive or only send. The ``sending'' nodes need not be real devices; they can be any acoustic events or signals of opportunity. We can distinguish the case when the sources and the receivers are synchronized so we can estimate the source--receiver distances, or the various cases when sources, receivers, or both sources and receivers are not synchronized.

\paragraph{Known Source Emission Times and Receiver Offsets} 
If the emission times happen to be known and the nodes and events all have a common time reference (that is, they are synchronized)\footnote{Additionally, the internal delays of the receivers are known.}, then the TOAs correspond to times of flight and directly give the distances between the nodes and the events. Given these distances, the joint localization problem reduces to multidimensional unfolding \cite{Schonemann:1970wd, Crocco:2012eu}. Some methods are based on direct optimization of the maximum likelihood (ML) criterion \cite{Weiss:1989array}, but these often fail due to the non-convexity of the objective and the existence of bad local minima \cite{wang2015self}. Other methods are based on Euclidean distance matrices and semidefinite programming (SDP) \cite{dokmanic2015relax}. 

Crocco et al. \cite{Crocco:2012eu} proceed by constructing a certain low-rank matrix of differences of squared TOAs from which the positions can be recovered by solving a low-dimensional non-convex optimization problem as follows. The low-rank matrix of differences of squared TOAs \cite{Crocco:2012eu} is related to the (translated) source positions $\mS$ and receiver positions $\mR$ as $\mH \approx \bar{\mR}^T \bar{\mS}$, where the latter is the matrix of inner products between \emph{centered} receiver points and \emph{centered} source points. The estimated $\mH$ is decomposed using singular value decomposition as $\mH = \mU \Sigma \mV^T$. Up to a translation due to centering, the source and receiver locations are then $\bar{\mS} =  \mQ^{-1}\Sigma \mV^T$ and $\bar{\mR} = (\mU\mQ)^T$ for some unknown invertible $3 \times 3$ matrix $\mQ$. Since we have that $\mH = \bar{\mR}^T \mQ \mQ^{-1} \bar{\mS}$, finding the correct relative geometry (between $\mR$ and $\mS$) amounts to identifying the right $\mQ$ and the difference vector between the center of $\mR$ and the center of $\mS$ which is done via non-convex optimization and can suffer from local optima, especially in the presence of noise. In our approach, we avoid the need to estimate $\mQ$ to \emph{stitch} $\mR$ and $\mS$ together by working with the full point set $\mX = [ \mR ; \mS]$ and the corresponding Gram matrix, instead of splitting $\mX$ into $\mR$ and $\mS$. The Gram matrix and its constraints then automatically encode the relative geometry and \emph{glue} together the sources and the receivers (see Section \ref{sub:characterization}).

\paragraph{When One Set of Times is Known} A more realistic scenario is when the source emission times are unknown and different but the microphones are synchronized. The reverse case is also relevant: when the receivers are not synchronized but the source emission times are known. This is the case if we use a distributed sensor array. 

The two unknowns (geometry and one set of times) can be estimated by directly optimizing an objective, for example using majorization
\cite{ono2009blind}. Again, this often fails due to the non-convexity of the problem and cannot work with near-minimal configurations. 

An alternative approach is to estimate the unknowns sequentially, first recovering the unknown times, and then using them to recover the geometry \cite{zhang2016iwaenc,Gaubitch:2013km}. Early work of Pollefeys and Nister \cite{pollefeys2008direct} exploits the low rank of a certain matrix of squared TOA differences. Their work is a near-field generalization of the work of Thrun \cite{thrun2006affine}, which was also adapted to work with missing measurements \cite{krekovic2018structure}. Heusdens and Gaubitch propose a more robust scheme based on structured total-least-squares \cite{Heusdens:2014er} to reconstruct the times.\footnote{Heusdens and Gaubitch in fact address the case when the microphone delays are unknown and the source is periodic; this is equivalent to only having unknown microphone delays; see Section \ref{sub:constantoffset}.} 

\paragraph{Unknown Emission Times and Receiver Offsets} The closest to our work is the method of Wang et al. \cite{wang2015self} who treat the most general case with unknown source emission times and receiver offsets. Same as the methods that address the case of one unknown set of offsets \cite{Heusdens:2014er}, Wang et al. proceed in two stages. The timing information is estimated via structured least-squares (Gauss--Newton), noting that with the correct timing estimates a certain matrix computed from the measured T(D)OA measurements must become low rank. This timing estimation procedure may require numerous random restarts. Once the timings are estimated, they proceed as Crocco et al. \cite{Crocco:2012eu} to recover the geometry. 

\paragraph{Convex Relaxations}
In the context of sensor network localization (with fixed anchor nodes), Biswas et al.~\cite{biswas2006semidefinite1,biswas2006semidefinite2} proposed to use SDP relaxations \cite{alfakih1999, ding2010} to handle the non-convexity introduced by the square root in the distance measurements. Similar relaxations have since also been used for source node localization from TDOA measurements (see e.g.~Yang et al.~\cite{yang2009efficient}, Vaghefi and Beuhrer~\cite{vaghefi2013asynchronous}) and mixed TDOA/FDOA measurements (Wang et al.~\cite{wang2012semidefinite}). Yang et al. \cite{yang2009efficient} also proposed a SOCP relaxation; however, it requires that the target node lies within the convex hull of the anchors. Jiang et al.~\cite{jiang2013time} propose to use Truncated Nuclear Norm Regularization (TNNR) to solve the TDOA self-calibration problem. Their optimization scheme consists of solving a sequence of convex surrogate problems based on the (convex) nuclear norm. We propose SDP relaxations for the full self-calibration problem, where both senders are receivers are unknown and not synchronized.

\paragraph{Minimal Estimation Problems}
A series of work considers minimal problems in TOA, where the goal is to estimate the unknowns from as few measurements as possible. Following up on their work on minimal problems for TOA measurements~\cite{kuang2013complete}, Kuang et al.~\cite{kuang2013stratified} propose a stratified approach for estimating the unknown time offsets from TDOA measurements. Once the offsets are recovered, the solvers from their previous work \cite{kuang2013complete} can be applied to recover the sender and receiver positions. Zhayida et al.~\cite{zhayida2015toa} (and later Farmani et al. \cite{farmani2016tdoa}) considered the minimal solutions to the special case of dual microphone setups. Burgess et al.~\cite{burgess2015toa} proposed solutions for settings where either the sender or the receivers lie in a lower dimensional space.
Batstone et al.~\cite{batstone2019robust} consider the case of constant offset TDOA self-calibration (i.e.~transmissions have a known period but unknown offset); this is also a stratified approach which first solves for the unknown time offset~\cite{kuang2013stratified}. We show that our approach can also succeed in near-minimal configurations.



\subsection{Contributions} 
\label{sub:contributions}

Our work is motivated by the fact that the two-stage procedure is suboptimal. The (valid) reason to adopt sequential estimation are poor local minima in the loss function which involves both the times and the positions. From a statistical point of view, however, joint estimation is preferred. Further, timing estimation can fail or require many random restarts; the same holds for the non-convex optimization to recover the relative configurations from $\bar{\mR}^T \bar{\mS}$. The two-stage procedure also makes it challenging to exploit prior geometric information such us known distances or distance bounds.

In light of related work and the above discussions, we summarize our contributions as follows:
\begin{itemize}
    \item What we are usually interested in are the sensor positions. We thus formulate a new self-localization loss which is invariant to offsets and delays. The proposed loss uses non-squared distances as opposed to the usual squared; we prove that minimizing this objective yields the maximum likelihood estimate of the positions under Gaussian noise. It enables us to recover the geometry without worrying about the times and without random restarts.
    \item We formulate a non-convex semidefinite optimization problem in terms of this new loss. We work with the full point set $\mX = [ \mR ; \mS]$ and the corresponding Gram matrix. The Gram matrix and its constraints intrinsically \emph{glue} together the sources and the receivers and thus obviate the need for the non-convex optimization step of Crocco et al. \cite{Crocco:2012eu} (see Section \ref{sub:characterization}); 
    \item Although our method does not require synchronization, it can  leverage synchronization among the receivers or the sources if it is fully or partially present \cite{Heusdens:2014er, Dokmanic2014tc,Dokmanic2016gu}. It can further leverage geometric side information such as when the receivers contain subarrays with known relative geometry or some sources and receivers are collocated. Finally, it can handle missing measurements.
\end{itemize}

The proposed method achieves state-of-the art results, not only in terms of localization accuracy, runtime, or exploiting side information, but also in the ability to solve near-minimal problems with few nodes. Finally, we note that as suggested by Ono \cite{ono2009blind}, our method can also be interpreted as a virtual synchronization method.



\section{Problem Statement} 
\label{sec:setup}

We wish to localize a set of $M$ receivers with unknown positions $\vr_1, \ldots, \vr_M \in \mathbb{R}^{d}$ using a set of $K$ sources with unknown positions $\vs_1, \ldots, \vs_K \in \mathbb{R}^{d}$. We assume that the sources emit pulses whose times of arrival (TOA) can be measured at the receivers.%
\footnote{Note that these measurements can be obtained without physically emitting pulses. In audio, for example, one often emits long chirps which is followed by pulse compression.}
Since we do not assume synchronization between the sources and the receivers, the absolute emission time $\tau_k$ of the $k$th source is unknown. Since the sources are not necessarily synchronized among themselves, the differences $\tau_k - \tau_{k'}$ are also unknown. Similarly, since we do not assume synchronization among receivers (nor knowing their internal delays), the temporal frames of reference of each receiver are shifted by an unknown $\sigma_m$ with respect to some reference clock. The times of arrival of the $k$th source signal at the $m$th receiver thus become
\begin{equation}
    \label{eq:measured_toa}
    t_{mk} = v^{-1}\norm{\vr_m - \vs_k} + \sigma_m + \tau_k,
\end{equation}
where for simplicity we let both $\sigma_m$ and $\tau_k$ have arbitrary sign. We assume the transmission speed $v$ to be known and w.l.o.g. let $v=1$ for the remainder of the paper. 

Note that if we instead use the time-difference-of-arrival (TDOA), then with the first sensor as the reference, the TDOA is
\begin{align*}
\tilde{t}_{mk} &= t_{mk} - t_{1k}\\
& = \|\vr_m - \vs_k\| + (\sigma_m  - \sigma_1) - \|\vr_1 - \vs_k\| \\
&= \|\vr_m - \vs_k\| + \tilde{\sigma}_m+ \tilde{\tau}_k.
\end{align*}
And so we can still think of the TDOA $\tilde{t}_{mk}$ as a TOA but with modified emission times and offsets.

\subsection{Minimal Number of Sources and Receivers} 
\label{sub:minimal_number_of_sources_and_receivers}

Common sensor localization scenarios are either in the horizontal plane $(d = 2)$ or in 3D space ($d = 3$). The number of the degrees of freedom in the positions of sources and receivers is then $d(M + K)$. However, since rigid motions of the entire setup cannot be distinguished from TOA data, we can choose the associated $d(d + 1)/2$ degrees of freedom freely ($d$ translational and $\binom{d}{2}$ rotational). Each source and receiver come with an additional unknown time, which gives another $M + K$ degrees of freedom. However, we can choose a global time offset arbitrarily, which is another gauge of invariance and subtracts one degree of freedom. The total number of the degrees of freedom is thus
\[
    \#(\text{DOF}) = (d + 1)(M + K - \tfrac{d}{2}) - 1
\]

As measurements we get $MK$ TOAs. In order to get a dimension-zero solution set (a solution set that is a finite or countable set of points, but not necessarily unique), this number should at least match the number of the degrees of freedom. Solving for the minimal number of sources as a function of the number of receivers, we obtain the following relation
\[
    K \geq \frac{(d+1)M - d(d+1)/2 - 1}{M - (d + 1)}.
\]
For $d = 3$ this gives $K \geq (4M - 7)/(M - 4)$ implying that the smallest number of receivers that can be localized is $M = 5$ for which at least $K = 13$ sources are required to get a dimension-zero solution. Some other minimal cases are: $(M = 7, K = 7), (M = 13, K = 5)$. 

As we demonstrate with real and numerical experiments in Section \ref{sec:experiments}, our algorithms can operate in the vicinity of this minimal regime, where previous methods either fail or perform poorly.


\section{A Loss Insensitive to Offsets and Delays}
Instead of proceeding sequentially by first estimating the unknown times as in the case of the state-of-the-art methods \cite{Heusdens:2014er,wang2015self}, we note that the primary task is usually position estimation rather than timing estimation. Besides, when the positions are known, estimating the timings $\set{\sigma_m}_{m=1}^M$ and $\set{\tau_k}_{k=1}^K$  boils down to a simple algebraic exercise (see Section \ref{sub:estimating_offsets_and_delays}). We thus devise a data fidelity metric which is insensitive to the unknown timings, yet is only minimized with the correct positions.

We begin by writing \eqref{eq:measured_toa} in matrix form as
\begin{equation}
    \label{eq:measured_toa_mtx}
    \mT = \mDelta + \vsigma \vone_K^T + \vone_M \vtau^T,
\end{equation}
where $\mT \in \R^{M\times K}$, $\mDelta = (\delta_{mk})$ and $\delta_{mk} = \norm{\vr_m - \vs_k}$, $\vsigma = [ \sigma_1, \ldots, \sigma_M ]^T$ and $\vtau = [ \tau_1, \ldots, \tau_K ]^T$. Now we make the key observation: multiplying \eqref{eq:measured_toa_mtx} on both sides by a matrix which has an all-ones vector in the nullspace will annihilate the two terms that depend on the unknown times. 

Given an integer $L$, let $\mJ_L$ be a geometric centering matrix of size $L$, \footnote{We often indicate matrix and vector sizes in subscripts. While making the notation a bit clunkier, it helps keep track of dimensions.}
\begin{equation}
    \label{eq:geometric_centering_mtx}
    \mJ_L := \mI_L - \tfrac{1}{L} \vone_L \vone_L^T.
\end{equation}
It is easy to verify that $\mJ_M \vone_M = \vzero_M$ and $\vone_K^T \mJ_K = \vzero_K^T$ so that the matrix
\begin{equation}
    \label{eq:P-matrix}
    \mP = \mJ_M \mT \mJ_K = \mJ_M \mDelta \mJ_K
\end{equation}
does not depend on the unknown times. In fact, we can say more.
\begin{proposition}
    \label{prop:cost_equivalence}
    We have $\mJ_M \mT_1 \mJ_K = \mJ_M \mT_2 \mJ_K$ if and only if there exist $\vsigma'$ and $\vtau'$ such that $\mT_1 = \mT_2 + \vsigma' \vone_K^T + \vone_M \vtau'^T$. In particular, for $\mT$ as in \eqref{eq:measured_toa_mtx}, we have $\mJ_M \mT \mJ_K = \mJ_M \mDelta \mJ_K$ \eqref{eq:P-matrix}.
\end{proposition}
\emph{Remark:} If $\mDelta$ happened to be a Euclidean (squared) distance matrix (EDM) of a point set $\mX$, for example $\mD = (d_{ij}^2)$, $d_{ij} = \norm{\vx_i - \vx_j}$, then $-\tfrac{1}{2} \mJ_N \mD \mJ_N$ would equal the Gram matrix of the (centered) $\mX$. No similar statement can be made here simply because $\mDelta$ holds non-squared distances, and because it only corresponds to one off-diagonal block of $\mD$. 

\begin{proof}
    The nullspace of the linear operator 
    \begin{align*}
        \calJ : \R^{M \times K} &\to \R^{M \times K} \\ 
        \mA &\stackrel{\calJ}{\mapsto} \mJ_M \mA \mJ_K
    \end{align*}
    is spanned by
    \[
        \calN(\calJ) = \mathrm{span} \{ \vone_M \ve_1^\T, \ldots, \vone_M \ve_K^T, \ve_1 \vone_K^T, \ldots \ve_M \vone_K^T \}.
    \]
    Note that not all of those $MK$ matrices are linearly independent, but the argument does not hinge on independence. Every matrix in the argument of $\mathrm{span}$ correspond to adding a constant offset to either a source or a receiver. As a consequence, any nullspace vector can be written as a linear combination of source / receiver offsets which proves the proposition.
\end{proof}

What Proposition \ref{prop:cost_equivalence} says is that $\mJ_M \mDelta \mJ_K$ is sufficient to determine the positions in the sense that the only locations such that this transformation matches the measurements are the true ones (if the original problem is solvable). 

We can then propose the following timing-invariant objective for localization

\begin{equation}
\underset{\mR,\mS}{\textrm{min}}\;  \frac{1}{2}  \norm{ \mJ_M (\mDelta(\mR,\mS)-\mT) \mJ_K }_F^2, \\
\label{eq:costxtdoa}
\end{equation}
where $\norm{\,\cdot\,}_F$ denotes the Frobenius norm, $\mR \in \R^{d\times M}$ are the receiver positions, and $\mS \in \R^{d\times K}$ are the source positions. This objective is justified by the following direct corollary of Proposition \ref{prop:cost_equivalence}.
\begin{corollary}
    If the unsynchronized localization problem has a unique solution (up to a rigid transformation), then vanishing loss in \eqref{eq:costxtdoa} implies correct localization.
\end{corollary}

Further, the loss  \eqref{eq:costxtdoa} is in fact equivalent to the ML loss \textit{given the offsets} $\vsigma$ and $\vtau$. Namely, assuming i.i.d. Gaussian noise on the TOA measurements we have
\[
    f_{ML}(\mDelta,\vsigma,\vtau) = \| \mDelta + \vsigma \vone_K^T + \vone_M \vtau^T - \mT \|_F^2,
\]
and the loss in \eqref{eq:costxtdoa} is simply
\[
    f(\mDelta) = \min_{\vsigma, \vtau} f_{ML}(\mDelta,\vsigma,\vtau).
\]
To see this note that
\begin{align}
    &\nabla_{\vsigma} f_{ML} = 2( \mDelta + \vsigma \vone_K^T + \vone_M \vtau^T - \mT )\vone_K = 0 \\ 
    \implies & \vsigma^\star = -\frac1{\vone_K^T\vone_K} ( \mDelta + \vone_M \vtau^T - \mT )\vone_K.
\end{align}
Inserting into the original cost we get
\[
   f_{ML}(\mDelta, \vsigma^\star, \vtau) = \| (\mDelta + \vone_M \vtau^T - T)\underbrace{(I_K - \frac1{K}\vone_K\vone_K^T)}_{J_K} \|_F^2.
\]
Solving for $\vtau^\star$ similarly yields $f_{ML}(\mDelta, \vsigma^\star, \vtau^\star) = f(\mDelta)$.

\subsection{Characterization in Terms of the Full Gram Matrix} 
\label{sub:characterization}
As described in Section \ref{sub:related_work}, a number of state-of-the-art methods based on low-rank matrix decompositions recover the matrix $\mH = \bar{\mR}^T \bar{\mS}$, with $\bar{\mR}$ and $\bar{\mS}$ being some translated (centered) versions of $\mR$ and $\mS$ \cite{Crocco:2012eu,Heusdens:2014er,wang2015self}. Matrix factorizations such as singular value decomposition can only determine $\bar{\mR}$ and $\bar{\mS}$ up to a multiplication by an arbitrary invertible matrix $\mQ$, since for any such $\mQ$ we have that $\mH = \wh{\mR}^T \wh{\mS}$, with $\wh{\mR} = \mQ^T\mR$, $\wh{\mS} = \mQ^{-1} \mS$. Finding the right $\mQ$  (which can be assumed to be upper-triangular which fixes the rotational gauge freedom) and the (one) translation vector is then achieved via direct non-convex optimization, though over a lower dimensional space than the original problem. 

We sidestep this problem of having to determine $\mQ$ in order to \emph{stitch} together $\mR$ and $\mS$ by writing the measurements in terms of the full point set 
\[
    \mX := [ \mR, \ \mS] \in \R^{d \times N},
\]
where $N = M+K$, and always working with the entire $\mX$. We denote the corresponding full Gram matrix by $\mG = \mX^T \mX \in \R^{N \times N}$. The off-diagonal blocks of the Gram matrix $\mG$ encode the relative geometry between $\mR$ and $\mS$---by recovering $\mG$, we know that relative geometry. Note that this only resolves the relative ambiguity between $\mR$ and $\mS$ which is
an artifact of the previous methods. The inherent invariance of localization from distances to Euclidean motions remains.

The matrix of squared distances (the EDM) between the column vectors in $\mX$ can be written as a linear function of the Gram matrix. Letting $\mD = (\norm{\vx_n - \vx_{n'}}^2)_{n,n'=1}^N$, one has
\begin{equation}
    \mD = \calD(\mG) := \diag(\mG) \vone_N^T - 2\mG + \vone_N \diag(\mG),
\end{equation}
with $\diag(\mG) = [ g_{11}, \ldots, g_{NN}]^T$. This means that the matrix of distances $\mDelta$ between sources and receivers can be written as an entrywise square root of a linear function of $\mG$, since
\begin{equation}
    \mDelta^{\bullet 2} = \calL(\mG) \bydef \mS_{\text{row}} \, \calD(\mG) \, \mS_{\text{col}},
\end{equation}
where $\mS_{\text{row}} := [\mI_M,\ \mat{0}_{M \times K}]$ and $\mS_{\text{col}} := [\mat{0}_{K\times M},\ \mI_K]^T$ are the appropriate row and column selection matrices, and $(\,\cdot\,)^{\bullet 2}$ denotes an entrywise square. In view of Proposition \ref{prop:cost_equivalence}, this leads us to the first reformulation of the original problem:
\begin{subequations}
\begin{alignat}{2}
&\!\min_{\mG}        &\qquad& \norm{\mJ_M (\sqrt[\bullet]{\calL(\mG)} - \mT) \mJ_K}_F^2 \label{eq:objective_noncvx}\\
&\text{subject to} &      & \mG \succeq 0,\\
&                  &      & \mG \vone_N = \vzero, \label{eq:centering1}\\
&                  &      & \rank(\mG) = d. \label{eq:rank_constraint}
\end{alignat}
\end{subequations}
The constraint \eqref{eq:centering1} resolves the translation ambiguity; it is equivalent to the point set being centered around the origin $\frac{1}{N} \sum_n \vx_n = 0$. 

Note that now the receivers and sources are being translated together, unlike in earlier work where they were split. Further, the estimated locations are directly read out from a factorization of $\mG$, without the need for stitching. 

\subsection{Semidefinite Relaxation}
The formulation \eqref{eq:objective_noncvx} remains nonconvex. Thus, direct approaches such as randomly-initialized first-order methods (e.g., projected gradient descent) are likely to get stuck in a local minimum. We thus propose to instead solve a convex relaxation of the problem.

There are two sources of nonconvexity: the entrywise square root in the objective, and the rank constraint on the Gram matrix. We begin by replacing the entrywise square root $\sqrt[\bullet]{\calL(\mG)}$ by a new matrix variable $\mB$, and adding the constraint that the entrywise square of $\mB$ must equal $\calL(\mG)$.

Concretely, we write the objective as 
\[
    \min_{\mG, \mB} \ \norm{\mJ_M (\mB - \mT) \mJ_K}_F^2,
\]
and ask that $\mB^{\bullet 2} = \calL(\mG)$ to obtain an equivalent formulation. We now proceed by reformulating the added constraint as a rank constraint on some semidefinite matrix (or matrices). Since the entrywise square does not mix the entries of $\mB$, we can add such constraints on a per-entry basis. This is computationally more efficient (and empirically works better) than working with the entire vectorization of $\mB$ at once.

We can equivalently write the constraint $b_{mk}^2 = \calL(\mG)_{mn}$ as
\begin{align}
    &\mL_{mk} := 
    \begin{bmatrix}
        \calL(\mG)_{mk} & b_{mk} \\
        b_{mk} & 1
    \end{bmatrix}
    \succeq \vec{0} \\
    & b_{mk} \geq 0 \\
    & \rank(\mL_{mk}) = 1.
\end{align}

All the non-convexity is now lumped into the rank constraints, $\rank(\mG) = d$ and $\rank(\mL_{mk}) = 1$ for $1 \leq m, k \leq M, K$. The final step is to relax all of the rank constraints to get
\begin{subequations}
\begin{alignat}{2}
&\!\min_{\mG, \mB}        &\qquad& \norm{\mJ_M (\mB - \mT) \mJ_K}_F^2 \label{eq:objective_cvx} \\
&\text{subject to} &      & \mG \succeq 0,\\
&                  &      & \mG \vone_N = \vzero,\\
&                  &      &  \mL_{mk} \succeq \vec{0}, ~ \text{for all} ~ 1 \leq m, n \leq M, K,\\
&                   &     &  \mB \geq 0.
\end{alignat}
\label{eq:sdr}
\end{subequations}

We can finally reconstruct the point set from the recovered $\mG$ using singular value decomposition. We have $\mG = \mU\Lambda\mV^\T$, where $\Lambda = \diag(\sigma_1,\dots , \sigma_N)$ with the singular values $\sigma_i$ assumed to be sorted in decreasing order. We reconstruct the point set using $\hat{\mX} = [\diag(\sigma_1,\dots, \sigma_d),\vzero_{d\times (N-d)}]\mV^\T$. If the rank of the estimated $\mG$ is truly $d$, as it ideally should be since it describes a $d$-dimensional point set, then the trailing singular values would be zero anyway.

\subsection{Refinement by Levenberg-Marquardt} 

In general, solving the above semidefinite program gives point configurations which are good coarse approximations to the true geometry, although they are not good enough for most applications. The reason can be tracked down to the fact that after relaxing the rank constraints, the estimated matrices have higher rank than desired. Still, the positive semidefinite constraints constrain geometry sufficiently to get decent estimates in many situations. One strategy would be to employ various rank minimization strategies. Informally, we tried this and it works decently but it is very slow. 

Instead, we propose to refine the output of the SDR using the Levenberg-Marquardt (LM) algorithm. It is fast, accurate, and it works better. We have also empirically observed that simple gradient descent is \emph{much} slower requiring more iterations to converge.

To derive the LM updates, we need to compute the Jacobian of the loss in order to linearize it. The loss 
\[
    \calJ(\mR, \mS) = \norm{\mJ_M (\mDelta(\mR, \mS) - \mT) \mJ_K}_F^2,
\]
can be written as
\[
    \calJ(\mR, \mS) = \norm{f(\mR, \mS)}_2^2,
\]
where $f(\mR, \mS) = \mA \vdelta(\mR, \mS) - \vt$,
\[
    \mA = \mJ_K^T \otimes \mJ_M, \quad \vt = \mA \, \vectorize(\mT),
\]
and $\vdelta = \vectorize(\mDelta)$ with $\otimes$ denoting the Kronecker product. 
Since $\mDelta : \R^{d \times (M + K)} \to \R^{M \times K}$, the Jacobian is a tensor in $\R^{M \times K \times d \times (M+K)}$. We can compute
\[
    [\mathsf{D} \mDelta]_{m, k, :, \ell} = 
    \begin{cases}
        \dfrac{\vr_m - \vs_k}{\norm{\vr_m - \vs_k}} & \ell = m, \\
        -\dfrac{\vr_m - \vs_k}{\norm{\vr_m - \vs_k}} & \ell = M + k, \\
        0 & \text{otherwise}.
    \end{cases}
\]
with the understanding that the first $M$ slices in the last index of $\mathsf{D}$ correspond to the receivers and the last $K$ to the sources. Note that the Jacobian is not well defined at $\vr_m = \vs_k$. In that case, it would be possible to remove the corresponding variables from the optimization and adjust the Jacobian accordingly.

To make things easier to compute with, we rewrite $\mT$ as a function of a single vector
\[
    \vx = \vx(\mR, \mS) =
    \begin{bmatrix}
        \vectorize(\mR) \\
        \vectorize(\mS)
    \end{bmatrix},
\]
and rearrange the output into a vector so that $\vdelta(\vx) = \vectorize(\mDelta(\mR, \mS))$. The Jacobian $\mathsf{D} \vdelta$ is obtained from $\mathsf{D} \mT$ by collapsing the first two dimensions and the last two dimensions so that $\mathsf{D} \vdelta \in \R^{MK \times d(M + K)}$. Finally the Jacobian of $f$ is computed as
\[
    \mathsf{D}f = \mA \ \mathsf{D} \vdelta,
\]
and we have an affine approximation of $f$ as (with some abuse of notation)
\[
    f(\vx; \vx_0) = f(\vx_0) + \mathsf{D}f(\vx)(\vx - \vx_0).
\]
This leads to the LM update,
\begin{align*}
    \vx^{(k+1)} = \argmin_{\vz} \ \norm{f(\vz;\vx^{(k)})}^2 + \lambda \norm{\vz - \vx^{(k)}}^2
\end{align*}
which is solved by
\[
    \vx^{(k+1)} = \vx^{(k)} - (\mathsf{D}f(\vx^{k})^T \mathsf{D}f(\vx^{k}) + \lambda \mI)^{-1} \mathsf{D}f(\vx^{k})^T f(\vx^{(k)}).
\]

The full localization algorithm is summarized in Algorithm \ref{alg:rec}.

\begin{algorithm}
\caption{Unsynchronized Sensor Localization.}
\label{alg:rec}
\begin{algorithmic}
\Require TOA $\mT \in \mathbb{R}^{M\times K}$
\Ensure Positions of receivers $\mR \in \mathbb{R}^{d\times M}$ and sources $\mS \in \mathbb{R}^{d\times K}$
\State Initialize $\mR$ and $\mS$ using SDR \eqref{eq:sdr}.
\State Update $\mR$ and $\mS$ using LM.
\end{algorithmic}
\end{algorithm}

We use the SeDuMi solver \cite{sturm1999sedumi} for the optimization problem \eqref{eq:sdr}. It has a $O(\sqrt{n}\log{\frac{1}{\epsilon}})$ worst-case time complexity per iteration where $n=d(M+K)$ is the number of variables. The LM refinement is $O(d^2(M+K)^2MK)$ per iteration.

\subsection{Estimating Offsets and Delays} 
\label{sub:estimating_offsets_and_delays}

Once an estimate $\wh{\mDelta}$ of $\mDelta$ is computed, we can also compute the unknown offsets $\vtau$ and delays $\vsigma$. We start by noting from \eqref{eq:measured_toa_mtx} that
\[
    \mT - \mDelta = \vsigma \vone_K^T + \vone_M \vtau^T,
\]
where the right hand side is linear in $(\vsigma, \vtau)$. We exploit it by vectorizing both sides:
\begin{align*}
    \ve := \vectorize(\mT - \mDelta) 
    &= \vone_K \otimes \vsigma + \tau \otimes \vone_M \\
    &= \mV \vsigma + \mW \vtau \\
    &= \begin{bmatrix} \mV & \mW \end{bmatrix} 
      \begin{bmatrix} \vsigma \\ \vtau \end{bmatrix},
\end{align*}
with $\mV := \vone_K \otimes \mI_M$, $\mW := \mI_K \otimes \vone_M$. From here, estimating the times vector amounts to solving an overdetermined system of linear equations. Note that the system matrix has a nullspace of dimension one (spanned by $\alpha\vone$) due to the global offset ambiguity. To eliminate the offset ambiguity we arbitrarily set $\sigma_1 = 0$, and let $\mV'$ be $\mV$ with the first column removed. The least-squares estimate of the unknown timings is then given as
\begin{equation}
    \begin{bmatrix} \wh{\vsigma}' \\ \wh{\vtau} \end{bmatrix} 
    = 
    \begin{bmatrix} \mV' & \mW \end{bmatrix}^\dagger \ve,
\end{equation}
with $\wh{\vsigma}' = [\wh{\sigma}_2, \ldots, \wh{\sigma}_M]^T$ and $( \, \cdot \, )^\dagger$ denoting the Moore-Penrose pseudoinverse.




\section{Variants} 

In many cases, we have some prior information about the distances or offset times. The proposed method makes it straightforward to leverage this prior knowledge. In the following, we explain how by describing several typical scenarios.

\subsection{Localization of Subarrays} 
\label{sub:localization_of_subarrays}

Suppose that some distances are known. It is often the case that the set of receivers consists of several compact subarrays with known geometry (voice-based assistants, smart phones). These compact subarrays are then distributed in the space of interest at unknown positions and with random orientations, together with discrete microphones.

Since the EDM is a linear function of the Gram matrix, this prior knowledge corresponds to simple linear constraints on $\mG$. We thus augment the original program with a set of \emph{known} inter-sensor distances for microphone pairs
\begin{align*}
    \calM 
    = \big\{&(m_1, m_2) \ : \ m_1 < m_2, \\
    &\text{distance}~d_{m_1 m_2}~\text{between}~\vr_{m_1}~\text{and}~\vr_{m_2}~\text{is known}\big\},
\end{align*}
Then we add the following set of constraints to our optimization:
\begin{equation}
    \calD(\mG)_{m_1 m_2} = d_{m_1 m_2}^2, 
    \ \text{for all} \
    (m_1, m_2) \in \calM, m_1 \neq m_2.
\end{equation}
If the set of receivers is partitioned as
\[
    \mR = [ \mR_1, \mR_2, \ldots, \mR_J],
\]
with the distances within the $j$th subarray $\mR_j$ known, these constraints correspond to knowing the diagonal subblocks of the upper-left block of $\calD$. 

We can again proceed with an LM refinement step. In this case, we consider the following objective
\begin{subequations}
\begin{alignat}{2}
&\!\min_{\theta}        &\qquad& \norm{f(\theta)}^2  \\
&\text{subject to} &      & g(\theta)=0,\\
\end{alignat}
\label{eq:knowndist}
\end{subequations}
where $g: \mathbb{R}^{d(M+K)} \mapsto \mathbb{R}^{|\calM|}$ is the known-distances residual defined as $$g_{m_1 m_2}(\mX) = \textrm{edm}(\mX,\mX)_{m_1 m_2} - d_{m_1 m_2}^2$$
with $\textrm{edm}(\mX,\mX)_{m_1 m_2} = \norm{\vx_{m_1} - \vx_{m_2}}^2$ being the entries of the EDM of the point set $\mX$. We solve \eqref{eq:knowndist} using the Augmented Lagrangian algorithm \cite{boydalgebra}. The corresponding update is given in Appendix \ref{app:subarrays}.

In some cases the distances may not be known exactly, but we might have access to good bounds. We once more leverage the linearity of $\calD$ in the Gramian by adding the following linear constraints to the program,
\begin{equation}
  \underline{d}_{m_1 m_2}^2 \leq \calD(\mG)_{m_1 m_2} \leq \bar{d}_{m_1 m_2}^2.
\end{equation}
where $\bar{d}_{m_1 m_2}$ and $\underline{d}_{m_1, m_2}$ are upper and lower bounds on the distance $d_{m_1, m_2}$.

Finally, we note that priors might be available not only for inter-receiver distances but also for inter-source distances and the distances between the sources and the receivers (we often have a coarse idea about how far the source events are from the microphones). These can be added to the constraints completely analogously. 


\subsection{When One Set of Times is Known} 
\label{sub:when_a_set_of_times_is_known}

When one set of timings is known (either the source emission times or the receiver offsets), it would still be possible to use the general formulation with the derived loss \eqref{eq:objective_noncvx} designed for the case when both sets of times are unknown \eqref{eq:measured_toa_mtx}. Though that would require more measurements than necessary given such prior knowledge as we show next.

Both cases (unknown emission times or unknown offsets) are captured by the simplified measurement model
\begin{equation}
    \label{eq:measured_toa_mtx_known_times}
    \mT = \mDelta + \vone_M \vtau^T,
\end{equation}
up to a transposition as necessary (exchanging the role of sources and receivers). From here, it follows that the influence of the timings can be eliminated simply by one left multiplication by $\mJ_M$ so that 
\[
    \mP' = \mJ_M \mT = \mJ_M \mDelta 
\]
does not depend on $\vtau$. We can thus replace the loss \eqref{eq:objective_noncvx} by 
\begin{equation}
    \norm{\mJ_M (\mB - \mT)}_F^2.
\end{equation}

To see how this reduces the minimal required number of measurements (sources or sensors), recall the degrees-of-freedom count from Section \ref{sub:minimal_number_of_sources_and_receivers}.  Concretely,
                \[
    \#(\text{DOF}) = d(M + K - \tfrac{d+1}{2}) + M
\]
gives
\[
    K \geq \frac{(d + 1)(2M - d)}{2(M - d)}.
\]
For $d = 3$, this becomes
\(
    K \geq \frac{4M - 6}{M - 3},
\)
so that the minimal number of receivers is now $M = 4$ and the minimal cases for the dimension-zero solution are $M = 4, K = 10$, $M = 5, K = 7$, $M = 6, K = 6$, and $M = 9, K = 5$. Another way to see this is by noting that the operator $\mA \mapsto \mJ_M \mA$ has a smaller nullspace than $\mA \mapsto \mJ_M \mA \mJ_K$.

\subsection{Sources with Known (Constant) Offset} 
\label{sub:constantoffset}

As a last example of prior information that is easily handled, we mention the practically relevant case where the source is for example a smartphone emitting periodic pulses. More generally, the source is emitting signals at known (not necessarily regular) intervals. The phone is not synchronized with the receivers, so the emitting times are strictly speaking unknown, but only up to one unknown offset---the start time of the emissions. This case has been studied in previous work \cite{Heusdens:2014er,batstone2019robust}.

We show how to incorporate this using our proposed method. We have that
\[
    \tau_k = \tau_o + \delta_k,
\]
with $\delta_k$ is the known offset of the $k$th emission with respect to the unknown time $\tau_o$. When the emissions are periodic, we have $\delta_k = k \cdot \delta$, but for simplicity we keep the notation more general.

Note that in general, the vectors $\vsigma$ and $\vtau$ can never be recovered uniquely, unless one of the times is known. The reason for this is that the global time reference can be decided arbitrarily (we used this fact when counting the degrees of freedom in Section \ref{sub:minimal_number_of_sources_and_receivers}). 

It then follows that the case of sources with known offsets is mathematically equivalent to the case when only the receivers' offsets are unknown. Concretely, first write
\[  
    \vtau = \tau_o \vone_K + \vdelta,
\]
where $\vdelta$ is a known vector. Then note that
\[
    \vsigma \vone_K^T + \vone_M \vtau^T 
    = (\vsigma + \tau_o \vone_M) \vone_K^T + \vone_M \vdelta^\T.
\]
Since $\vone \vdelta^\T$ is known it can be absorbed in the measurements; the one unknown offset $\tau_o$ can then be absorbed in receiver offsets.


\subsection{Missing Measurements} 
\label{sub:missing_measurements_and_robust_estimation}

One upshot of our formulation is that it allows localization even when some sensor--receiver measurements are unavailable. We denote the set of missing entries by 
\begin{equation*}
\begin{aligned}
 \calM = \big\{(m, k) :{} & \text{distance between}~m^{\textrm{th}}~\text{receiver and }\\
 & k^{\textrm{th}}~\text{source is missing}\big\}.\\
\end{aligned}
\end{equation*}

We then replace the objective \eqref{eq:objective_cvx} with
\begin{equation}
\min_{\mG, \mB, \set{\alpha_{mk}}} \  \bigg\|\mJ_M (\mB - \mT + \sum_{\mathclap{(m, k) \in \calM}} \alpha_{mk} \ve_m \ve_k^\T) \mJ_K\bigg\|_F^2, 
\end{equation}
where $\ve_n$ denotes the canonical basis, and the coefficients $\alpha_{mk}$ account for the missing entries.

The LM refinement then proceeds with $\theta = \begin{bmatrix}
\vectorize(\mR)\\
\vectorize(\mS)\\
\vectorize(\boldsymbol{\alpha})
\end{bmatrix} \in \R^{d(M+K)+|\calM|}$. The Jacobian now includes the derivative with respect to $\boldsymbol{\alpha}$ as well
\begin{equation}
[\mathsf{D}_{\boldsymbol{\alpha}}f]_{mk, ij} = 
    \begin{cases}
        1 & (m,k) \in \calM, \; i=m, j=k, \\
        0 & \text{otherwise}.
    \end{cases}
\end{equation}

Note that this approach can be used in the presence of outliers by first estimating them using RANSAC for instance. Then those outliers can be treated as missing entries.


\section{Numerical Simulations}
\label{sec:numerical}
In this section, we evaluate our approach on synthetic data. We implemented our algorithms in Matlab and used the CVX package for specifying convex programs \cite{cvx,gb08} with the SeDuMi solver. For reverberation simulations, we use the Pyroomacoustics~\cite{scheibler2018} package in Python. In the following, we describe the setup and present a series of results for different scenarios showcasing the accuracy and  flexibility of our method.

\subsection{Setup}
We consider an equal number of sensors and sources $M=K$ in the range of 7 to 12. For each pair, we generated 200 configurations of points randomly distributed in a volume of dimensions 10 m $\times$ 10 m $\times$ 3 m. The time offsets are uniformly generated in the range $[-1,1]$ s. The TOAs were finally corrupted by zero-mean random Gaussian noise with standard deviation $\sigma \in \{10^{-3},10^{-4},10^{-5},10^{-6,},0\}$ s or equivalently $\sigma \in \{34.3,3.4,0.3,0.03,0\}$ cm. The parameters are summarized in Table \ref{tab:xtdoaparam}.

\begin{table}[!htb]
\caption{Parameters for the setup and algorithms.}
\label{tab:xtdoaparam}
\centering
\begin{tabular}{ll}
\toprule
Dimension & $d=3$\\
Volume &  10 m $\times$ 10 m $\times $ 3 m \\
M sensors  & 7 to 12 \\
K sources  & 7 to 12\\
Time offset & [-1,1] s \\
Speed of sound & 343 m/s\\
Solver & SeDuMi \\
LM iterations & At most 1000\\
\bottomrule
\end{tabular}
\end{table}

\subsection{Evaluation}

The reconstructed points are first optimally aligned using Procrustes analysis \cite{schoenemann1966}. The localization error between the true $[\mR,\mS]$ and reconstructed $[\hat{\mR},\hat{\mS}]$ points is then
\begin{equation}
E_{\textrm{rs}} = \frac{\sum \limits_{m=1}^{M} \norm{\vr_m - \hat{\vr}_m} + \sum \limits_{k=1}^{K} \norm{\vs_k - \hat{\vs}_k}}{M+K}.
\end{equation}
We present the results using box plots depicting the median, first quartile, and third quartile. The whiskers correspond to 1.5 times the inter-quartile range. Values over or below this are shown as outliers. 

To facilitate comparison, we will also present the median and the 95\% confidence interval. In all figures, we clip the errors to a minimum of $10^{-3}$ which we assume is sufficient for most applications.

\subsection{Results}
As a baseline algorithm, we choose to compare to the two-stage approach of Wang et al. \cite{wang2015self}. We consider two aspects in our comparison: localization error and runtime. Note that we only report the results of Wang et al.'s approach without the further third-stage refinement using Ono et al.'s \cite{ono2009blind} algorithm. We had observed that the refinement takes a long time due to the required number of iterations and did not improve the results in the case $M=K=7$.

\begin{figure*}
\centering
\begin{minipage}[b]{0.75\linewidth}
  \centering
 \centerline{\includegraphics[width=\linewidth]{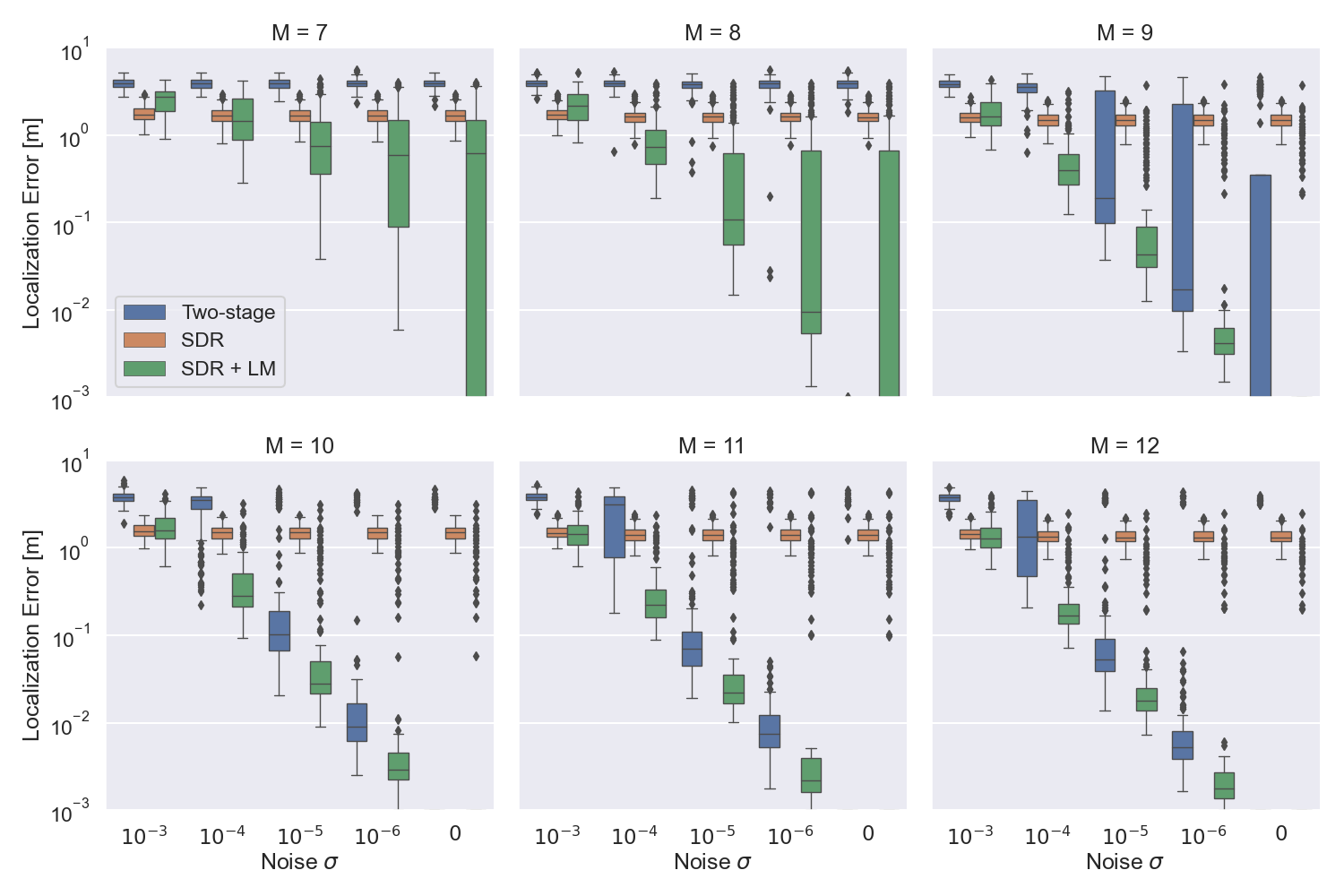}}
 \medskip
\end{minipage}
\caption{Localization results for an unsynchronized setup comparing our approach (SDR+LM) to the two-stage baseline (Wang et al.~\cite{wang2015self}). We also show the results before the LM refinement (SDR). Results are shown at different noise levels and for $M=K$ in the range of 7 to 12.}
\label{fig:comparewang}
\hfill
\end{figure*}

\begin{figure}
\centering
\begin{minipage}[b]{0.75\linewidth}
  \centering
   \centerline{\includegraphics[width=\linewidth]{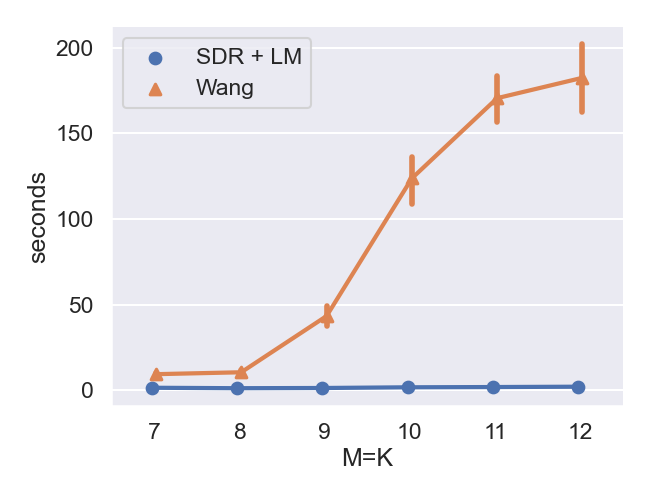}}
\end{minipage}
\caption{Average runtime of our approach (SDR+LM) compared to the two-stage baseline (Wang et al.~\cite{wang2015self}). Results are for $M=K$ in the range of 7 to 12. The noise standard deviation is fixed $\sigma=10^{-5}$. The runtime of our approach is consistently less than 5 seconds.}
\label{fig:runtime}
\hfill
\end{figure}

The localization results comparing our approach (SDR+LM) to the two-stage baseline are shown in Figure \ref{fig:comparewang}. We also show our SDR results before the LM refinement. Note that while the SDR-only results are poor, they indeed provide a good initialization for the subsequent LM refinement as evidenced by the SDR+LM results. We further note that the outlier errors correspond to geometrically unfavorable source/receiver configurations that have attractive local minima. As can be seen, our timing-invariant approach outperforms the two-stage method, especially in the near-minimal configurations. As for the runtime, we show in Figure \ref{fig:runtime} a comparison of the average runtime over 20 runs as we increase the number of sensors and receivers for a fixed noise standard deviation $\sigma=10^{-5}$. The runtime of our approach is consistently less than 5 seconds, whereas the runtime for the two-stage approach increases significantly with the number of sensors. Thus, our timing-invariant approach avoids both having timing-estimation errors that propagate to the position estimation as well as the need for multiple random initializations that increase the runtime. 

\begin{figure*}
\centering
\begin{minipage}[b]{0.75\linewidth}
  \centering
    \centerline{\includegraphics[width=\linewidth]{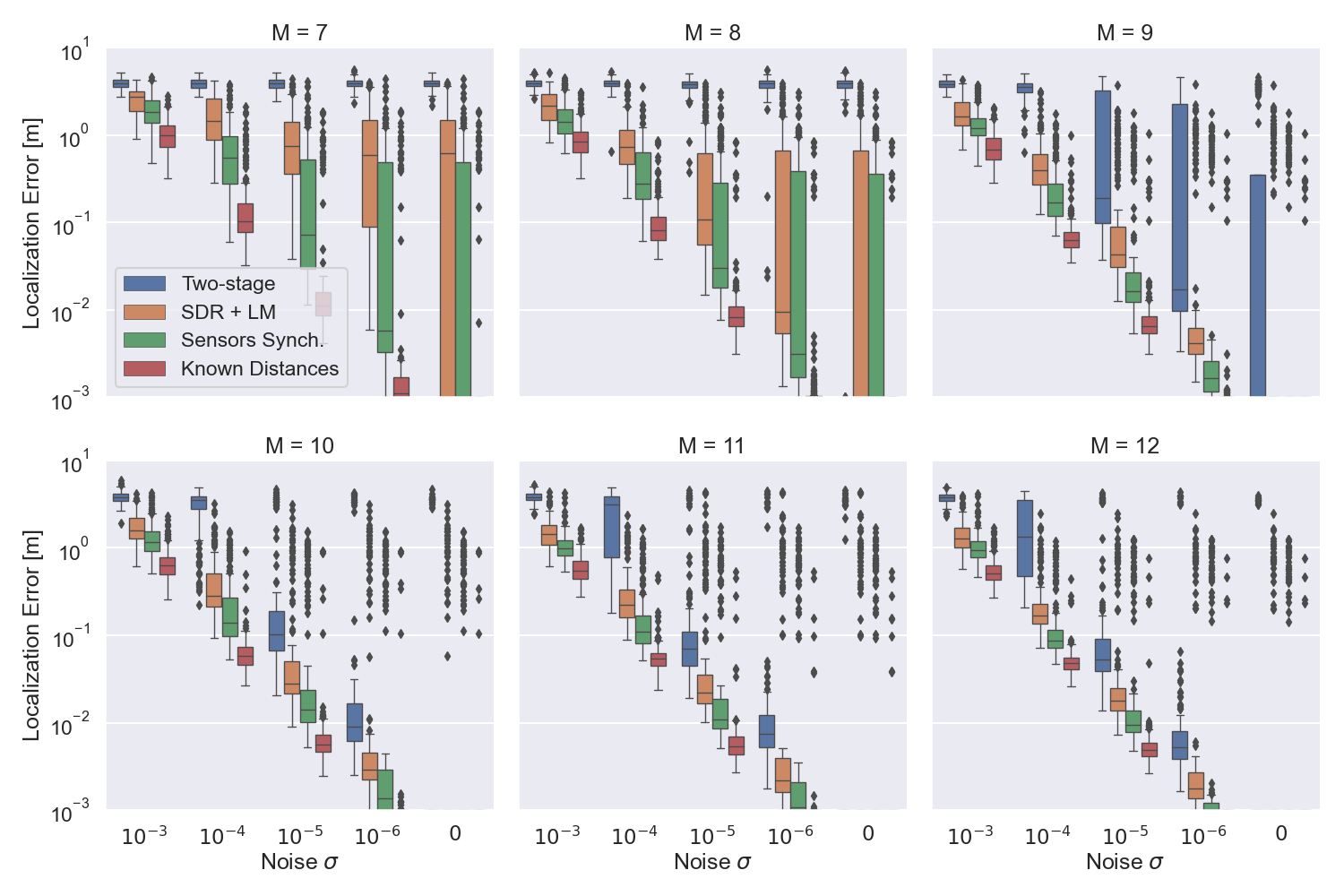}}
  \medskip
\end{minipage}
\caption{Localization errors for an unsynchronized setup in different scenarios. Results are shown at different noise levels and for $M=K$ in the range of 7 to 12. In the case without any synchronization, our approach (SDR+LM) outperforms the two-stage baseline (Wang et al.~\cite{wang2015self}). Side information such as partial synchronization or knowledge of some distances improves the results. }
\label{fig:compareall}
\hfill
\end{figure*}

We now evaluate the case when the microphones are synchronized but the source emission times are still unknown. The results are shown in Figure \ref{fig:compareall} along with the results of the fully unsynchronized case. As can be seen, the localization performance with partial synchronization is better, especially for $(M=7,K=7)$ and $(M=8,K=8)$. 

Next, we test the case when the inter-microphone distances are known, but the whole setup is still fully unsynchronized. The localization errors are shown in Figure \ref{fig:compareall}. As would be expected, having side information significantly improves the localization performance.

\begin{figure*}
\centering
\begin{minipage}[b]{0.75\linewidth}
  \centering
    \centerline{\includegraphics[width=\linewidth]{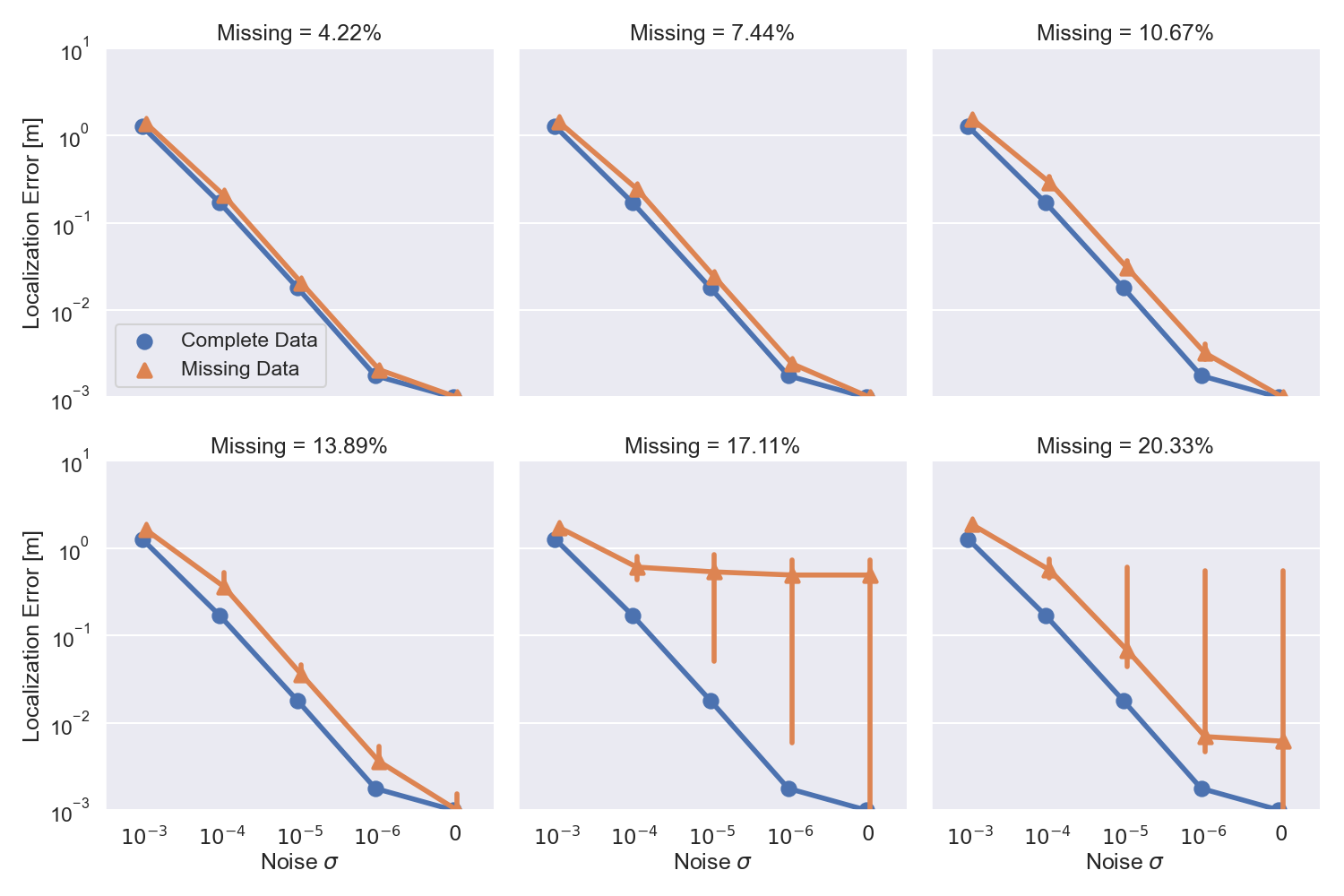}}
\medskip
\end{minipage}
\caption{(Missing Data.) Median localization errors for $M=K=12$ in the case of missing TOA entries at different levels of noise. Between 4\% and 20\% of entries are missing. The whole setup is unsynchronized. The results are plotted against those without any missing data. }
\label{fig:missing}
\hfill
\end{figure*}

For the last experiment, we attempt localization when some TOA data is missing. We test a range of missing entries from 4\% to 20\% of the total. Figure \ref{fig:missing} shows the results for $M=K=12$ where we also compare the performance to the complete data case. We can see that we are still able to localize when few entries are missing. However, as we have found some intrinsically bad configurations that cannot be accurately localized with complete data, it is expected that the problem would be exacerbated when some TOA entries are further missing. 

In summary, our approach (SDR+LM) outperforms the two-stage baseline (Wang et al.~\cite{wang2015self}) in terms of localization error and runtime. Additional side information such as partial synchronization or knowledge of some distances can be easily accommodated in our formulation, and as expected improve the localization results compared to the fully unsynchronized case.

\subsection{Reverberation Results}
\label{sub:reverb}
We now evaluate the localization performance of our approach in the presence of reverberation. 

We consider a room of dimensions $10\times10\times3$. The $M=K=12$ sensors and sources are randomly placed in the room. The sources are random Gaussian noise and do not overlap such that the segmentation at each microphone is known. We also test with speech sources from the Speech Commands dataset \cite{speechcommands}. The SNR is set to 15 dB.

We vary the reverberation time by adjusting the wall absorption coefficients and maximum order for the image-source method. For each reverberation time, we simulate 20 realizations. While the desired reverberation times are 0 s, 0.1 s, 0.2 s, 0.3 s, 0.4 s, and 0.5 s, the actual reverberation times obtained in simulation are slightly different but close. The parameters are summarized in Table \ref{tab:reverbparam}.

The TDOA are estimated using the method of Yamaoka et al.~\cite{yamaoka2019tdoa}. The error between the true $t_{mk}$ and estimated $\hat{t}_{mk}$ TDOA is calculated as 
\begin{equation}
E_{\textrm{t}} = \frac{\sum \limits_{m=1}^{M} \sum \limits_{k=1}^{K} |t_{mk} - \hat{t}_{mk}|}{MK}.
\end{equation}
We then use the estimated TDOA as input to our localization algorithm as well as the two-stage baseline of Wang et al.~\cite{wang2015self}.

\begin{table}[!htb]
\caption{Parameters for the reverberation simulation.}
\label{tab:reverbparam}
\centering
\begin{tabular}{ll}
\toprule
Room dimensions &  10 m $\times$ 10 m $\times $ 3 m \\
M=K sensors and sources  & 12 \\
Time offset & [0,2] s \\
Source duration & 0.5 s\\
Speed of sound & 343 m/s\\
Sampling rate & 48000 Hz for noise and 16000 Hz for speech\\
Reverberation time & 0 to 0.5 s \\
SNR & 15 dB\\
\bottomrule
\end{tabular}
\end{table}

\begin{figure}
\centering
\begin{minipage}[b]{0.75\linewidth}
  \centering
    \centerline{\includegraphics[width=\linewidth]{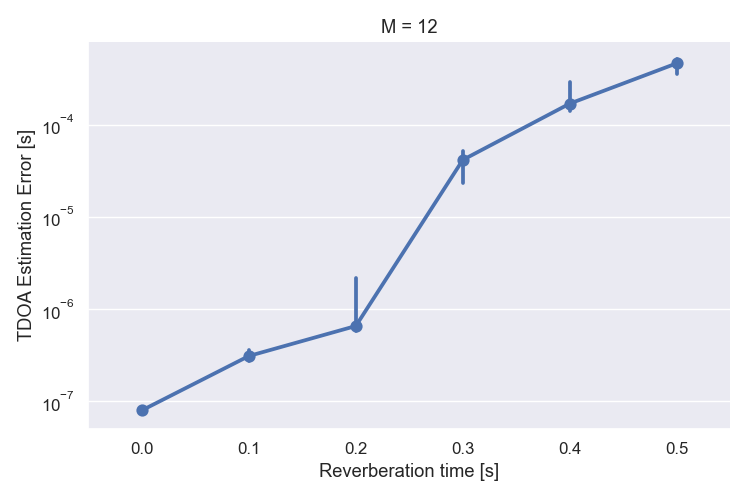}}
\centerline{(a)}\medskip
\medskip
\end{minipage}
\begin{minipage}[b]{0.75\linewidth}
  \centering
    \centerline{\includegraphics[width=\linewidth]{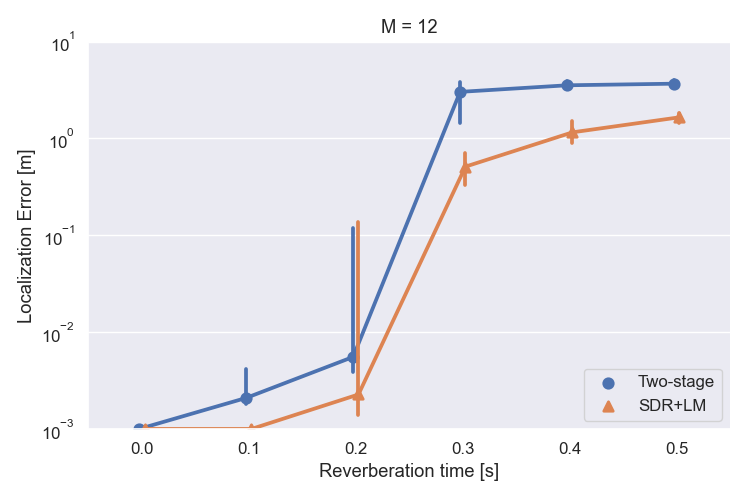}}
\centerline{(b)}\medskip
\medskip
\end{minipage}
\caption{(Reverberation.) Results with noise sources as the reverberation time increases. (a) Median TDOA estimation error. (b) Median localization error. Our timing-invariant approach outperforms the two-stage baseline.}
\label{fig:reverb}
\hfill
\end{figure}

\begin{figure}
\centering
\begin{minipage}[b]{0.75\linewidth}
  \centering
    \centerline{\includegraphics[width=\linewidth]{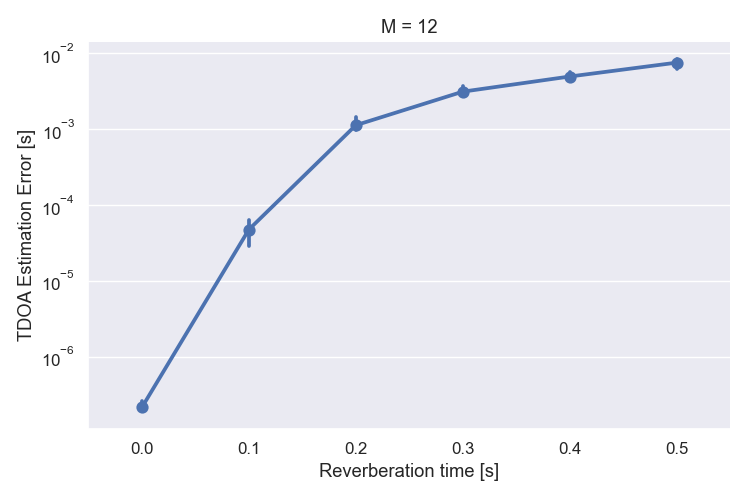}}
\centerline{(a)}\medskip
\medskip
\end{minipage}
\begin{minipage}[b]{0.75\linewidth}
  \centering
    \centerline{\includegraphics[width=\linewidth]{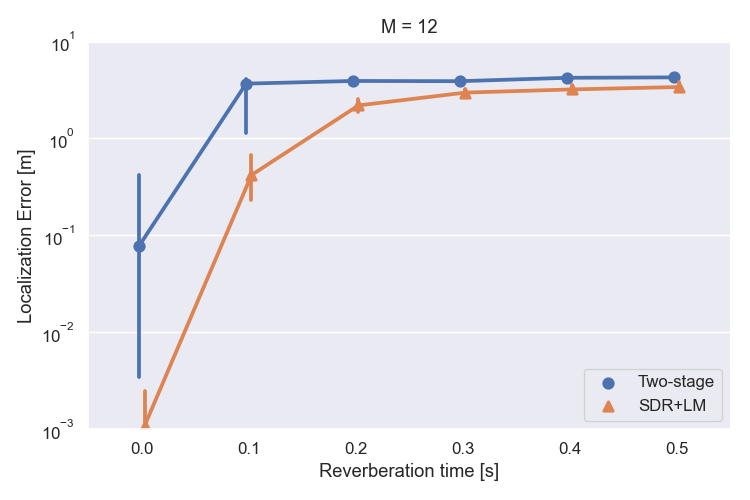}}
\centerline{(b)}\medskip
\medskip
\end{minipage}
\caption{(Reverberation.) Results with speech sources as the reverberation time increases. (a) Median TDOA estimation error. (b) Median localization error. The TDOA estimation errors are larger with speech compared to noise sources. The subsequent localization is also worse. Our timing-invariant approach outperforms the two-stage baseline.}
\label{fig:reverb_speech}
\hfill
\end{figure}

The results are shown in Figures \ref{fig:reverb} and \ref{fig:reverb_speech}. In Figure \ref{fig:reverb}a, we show the median TDOA estimation error. As expected, the errors are larger as the reverberation time increases. Subsequently, this affects the localization errors as shown in Figure \ref{fig:reverb}b. A similar trend is observed with the speech sources in Figure \ref{fig:reverb_speech}a. For both source types, our timing-invariant approach still outperforms the two-stage baseline.

\section{Experiments} 
\label{sec:experiments}
In this section, we evaluate our approach on real data \cite{batstone2019robust} recorded in an office with most of the furniture removed. A loudspeaker played a chirp from 65 positions. We have access to the ground truth positions for 12 microphones measured using a laser, the $12\times 65$ TOA matrix, and a mask indicating the positions of outlier TOA entries. While the TOA was extracted from the recordings knowing the chirp, our algorithms also work for TDOA matrices which could have been extracted without any assumption on the source as shown in Section~\ref{sub:reverb}. 

\subsection{Setup}
\begin{figure}
\centering
\begin{minipage}[b]{0.75\linewidth}
  \centering
 \centerline{\includegraphics[width=\linewidth]{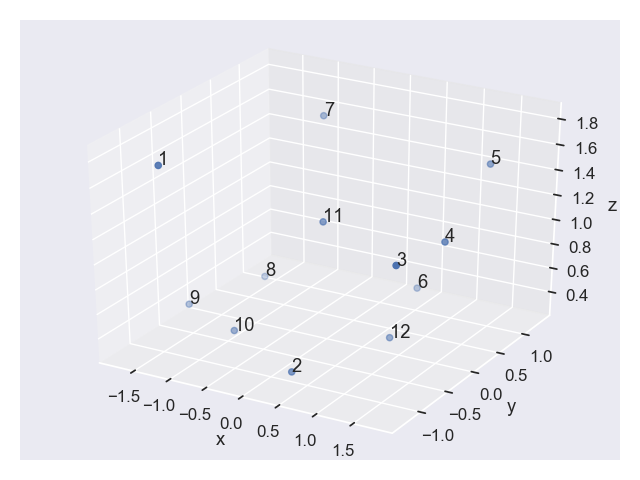}}
\centerline{(a)}\medskip
\end{minipage}
\begin{minipage}[b]{0.75\linewidth}
  \centering
 \centerline{\includegraphics[width=\linewidth]{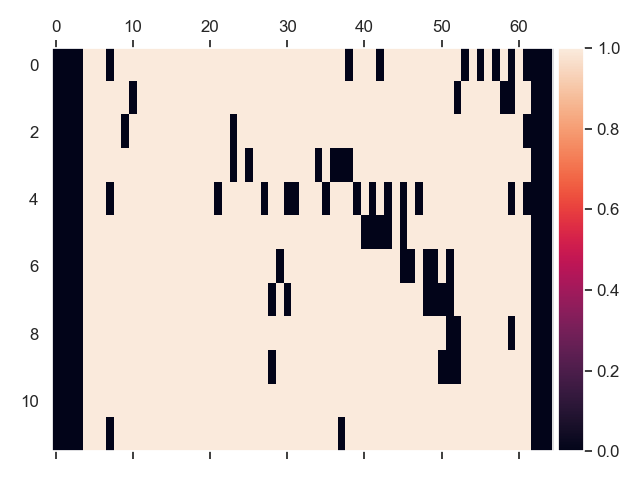}}
\centerline{(b)}\medskip
\end{minipage}%
\caption{(Real Data.) Data from a real experiment carried out in an office environment. (a) True positions of 12 microphones. (b) Mask showing outliers in the TOA data. }
\label{fig:realtoa}
\hfill
\end{figure}

Figure \ref{fig:realtoa} shows the TOA matrix, outlier positions, and the true microphone positions. A total of 23 loudspeakers provide clean data without outliers. Thus, for the first experiments, we will use the $12\times 23$ subset of TOAs. Then, for the full $12\times 65$ TOAs, we will use our missing data approach to handle the outliers. We report results over 200 runs where we randomly choose a subset of the loudspeakers to localize the 12 microphones.

Since we only have the ground truth for the microphone positions, we calculate the localization error between the true $\mR$ and reconstructed $\hat{\mR}$ microphone positions as
\begin{equation}
E_{\textrm{r}} = \frac{\sum \limits_{m=1}^{M} \norm{\vr_m - \hat{\vr}_m} }{M}.
\end{equation}

\subsection{Results}
\begin{figure}
\centering
\begin{minipage}[b]{0.75\linewidth}
  \centering
     \centerline{\includegraphics[width=\linewidth]{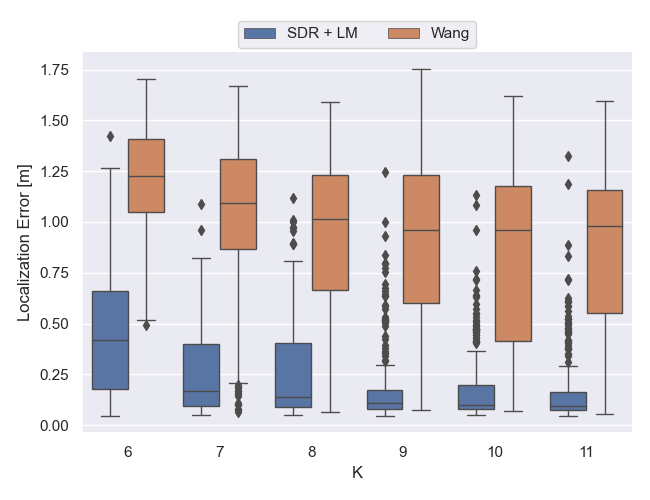}}
\end{minipage}
\caption{(Real Data.) Localization errors for localizing 12 microphones using 6 to 11 sources. We compare our approach (SDR+LM) to a two-stage approach (Wang). The minimum error with our approach is consistently less than 5 cm. The input TOA matrix is a subset of the $12\times 23$ outlier-free TOA matrix. }
\label{fig:realcomplete}
\hfill
\end{figure}

We first test our approach on subsets of the $12\times 23$ TOA matrix that is outlier-free. We also compare to the Wang et al. \cite{wang2015self} two-stage baseline. In Figure \ref{fig:realcomplete}(a), we show the errors for localizing all 12 microphones using a varying number of loudspeakers from 6 (the minimal case for $M=12$) to 11. Once again, our approach significantly outperforms the two-stage baseline. Also similar to what we observed in the numerical simulations, while using more loudspeakers improves the average error, for each $K$, there are a number of large errors corresponding to intrinsically difficult configurations that have attractive non-global minima. However, the minimum error is consistent for all $K$ and is less than 5 cm.

\begin{figure}
\centering
\begin{minipage}[b]{0.75\linewidth}
  \centering
  \centerline{\includegraphics[width=\linewidth]{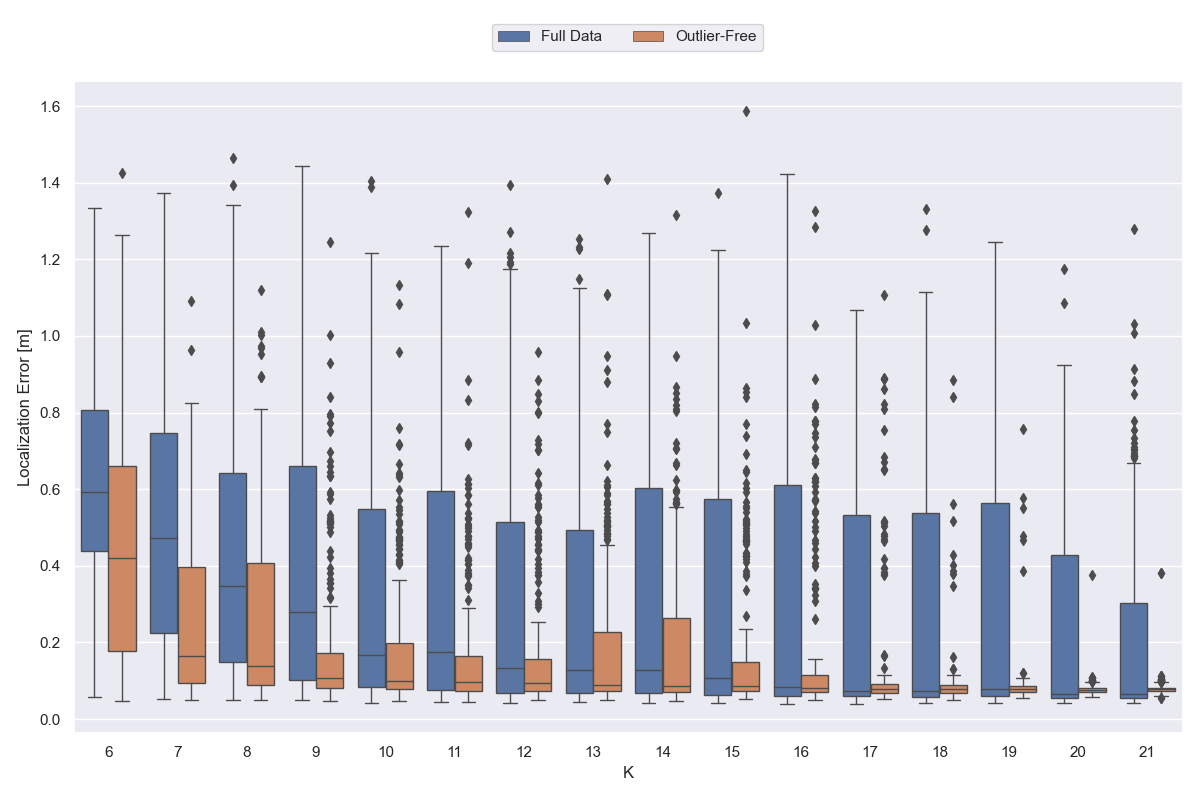}}
\end{minipage}
\caption{(Real Data.) Localization errors for localizing 12 microphones using 6 to 21 sources. We compare using subsets of the available $12\times 65$ TOA matrix (Full Data) to using subsets of the $12\times 23$ outlier-free TOA matrix (Outlier-Free). For the former, we use the missing data approach where we treat outliers as missing entries.}
\label{fig:realminimal}
\hfill
\end{figure}

Next, we use subsets of the entire $12\times 65$ TOA that contains outliers but assume knowing where the outliers are. We can thus use the missing data approach described in Section \ref{sub:missing_measurements_and_robust_estimation}.  Figure \ref{fig:realminimal} shows the errors for localizing the 12 microphones using 6 to 21 sources. On one hand, the average error is larger compared to the outlier-free case. On the other hand, the best case scenarios that result in minimum errors are comparable to the outlier-free case. The smallest localization error is 4 cm and corresponds to $K=16$ loudspeakers.

\section{Conclusion}
We formulated a timing-invariant objective that allows us to localize sensors and sources in an unsynchronized setting. Based on this objective, we proposed an SDP relaxation using the full Gram matrix of the point set and then a subsequent refinement step using the LM algorithm. We have thus eliminated the need for having to first estimate the unknown timings as well as avoided the multiple initializations required when solving non-convex problems. We also proposed an approach to handle missing data and showed how to seamlessly incorporate additional information such as knowledge of some distances or partial synchronization.

Using numerical simulations, we demonstrated the feasibility of the approach in different scenarios and in near-minimal configurations. We compared our approach to a two-stage state-of-the-art method \cite{wang2015self}. Not only did our approach outperform the two-stage algorithm in terms of localization error, but also in terms of runtime. We also tested our algorithms on real times of arrival measured in an office environment. We were able to localize the 12 microphones to within 0.04 m average error.

Future work could focus on dealing with outliers which arise in the presence of multipath, for example. The method could be similar to our missing data approach except that the positions of the outliers are not known and need to be determined.
\appendices

\section{Localization of Subarrays}
\label{app:subarrays}
The LM refinement proceeds by iteratively minimizing the augmented Lagrangian over $\theta$ 
\begin{equation}
 \norm{f(\theta)}^2 + g(\theta)^\T \vz + \mu \norm{g(\theta)}^2,
 \label{eq:aug}
\end{equation}
where $\vz$ is the Lagrange multiplier and $\mu>0$. Minimizing \eqref{eq:aug} is equivalent to minimizing \cite{boydalgebra}
\begin{equation}
 \norm{f(\theta)}^2 + \mu \norm{g(\theta) + \vz/(2\mu)}^2.
 \label{eq:aug2}
\end{equation}
Since \eqref{eq:aug2} is nonlinear in $\theta$, the LM algorithm is once again used in which the update is
\begin{alignat*}{1}
\theta^{(i+1)} =\; &\theta^{(i)} - \\ &\left(\mathsf{D}f(\theta^{(i)})^\T \mathsf{D}f(\theta^{(i)}) + \mu \mathsf{D}g(\theta^{(i)})^\T \mathsf{D}g(\theta^{(i)})+   \lambda^{(i)} \mI\right)^{-1} \\
 &\left(\mathsf{D}f(\theta^{i})^\T f(\theta^{(i)}) + \mu \mathsf{D}g(\theta^{(i)})^\T (g+\vz/(2\mu) \right)
\end{alignat*}
where $ \mathsf{D}g \in \R^{|\calM|\times d(M+K)}$ is the Jacobian defined as
\begin{equation}
\mathsf{D}g_{ij, p} = 
    \begin{cases}
       \vx_i - 2\vx_j & \; p=i,\\
       -2\vx_i + \vx_j &\; p=j, \\
        0 & \text{otherwise}.
    \end{cases}
\end{equation}

\bibliographystyle{IEEEtran}
\bibliography{xtdoa}

\begin{thebibliography}{10}
\providecommand{\url}[1]{#1}
\csname url@samestyle\endcsname
\providecommand{\newblock}{\relax}
\providecommand{\bibinfo}[2]{#2}
\providecommand{\BIBentrySTDinterwordspacing}{\spaceskip=0pt\relax}
\providecommand{\BIBentryALTinterwordstretchfactor}{4}
\providecommand{\BIBentryALTinterwordspacing}{\spaceskip=\fontdimen2\font plus
\BIBentryALTinterwordstretchfactor\fontdimen3\font minus
  \fontdimen4\font\relax}
\providecommand{\BIBforeignlanguage}[2]{{%
\expandafter\ifx\csname l@#1\endcsname\relax
\typeout{** WARNING: IEEEtran.bst: No hyphenation pattern has been}%
\typeout{** loaded for the language `#1'. Using the pattern for}%
\typeout{** the default language instead.}%
\else
\language=\csname l@#1\endcsname
\fi
#2}}
\providecommand{\BIBdecl}{\relax}
\BIBdecl

\bibitem{SanAntonio:2007mimo}
G.~San~Antonio, D.~R. Fuhrmann, and F.~C. Robey, ``{MIMO} radar ambiguity
  functions,'' \emph{IEEE Journal of Selected Topics in Signal Processing},
  vol.~1, no.~1, pp. 167--177, 2007.

\bibitem{Parhizkar:2013calibration}
R.~Parhizkar, A.~Karbasi, S.~Oh, and M.~Vetterli, ``Calibration using matrix
  completion with application to ultrasound tomography,'' \emph{IEEE
  transactions on signal processing}, vol.~61, no.~20, pp. 4923--4933, 2013.

\bibitem{ferguson1989underwater}
B.~G. {Ferguson}, ``Improved time-delay estimates of underwater acoustic
  signals using beamforming and prefiltering techniques,'' \emph{IEEE Journal
  of Oceanic Engineering}, vol.~14, no.~3, pp. 238--244, 1989.

\bibitem{devaney2005}
A.~Devaney, ``Time reversal imaging of obscured targets from multistatic
  data,'' \emph{IEEE Transactions on Antennas and Propagation}, vol.~53, no.~5,
  pp. 1600--1610, 2005.

\bibitem{ciuonzo2015}
D.~Ciuonzo, G.~Romano, and R.~Solimene, ``Performance analysis of time-reversal
  {MUSIC},'' \emph{IEEE Transactions on Signal Processing}, vol.~63, no.~10,
  pp. 2650--2662, 2015.

\bibitem{scheibler2018separake}
R.~Scheibler, D.~Di~Carlos, A.~Deleforge, and I.~Dokmani\'{c}, ``Separake:
  Source separation with a little help from echoes,'' in \emph{2018 IEEE
  International Conference on Acoustics, Speech and Signal Processing
  (ICASSP)}.\hskip 1em plus 0.5em minus 0.4em\relax IEEE, 2018, pp. 6897--6901.

\bibitem{pan2017frida}
H.~Pan, R.~Scheibler, E.~Bezzam, I.~Dokmani{\'c}, and M.~Vetterli, ``{FRIDA}:
  {FRI}-based {DOA} estimation for arbitrary array layouts,'' in \emph{2017
  IEEE International Conference on Acoustics, Speech and Signal Processing
  (ICASSP)}.\hskip 1em plus 0.5em minus 0.4em\relax IEEE, 2017, pp. 3186--3190.

\bibitem{Martinez:2013fukushima}
M.~Martinez-Camara, I.~Dokmani{\'c}, J.~Ranieri, R.~Scheibler, M.~Vetterli, and
  A.~Stohl, ``The fukushima inverse problem,'' in \emph{2013 IEEE International
  Conference on Acoustics, Speech and Signal Processing}.\hskip 1em plus 0.5em
  minus 0.4em\relax IEEE, 2013, pp. 4330--4334.

\bibitem{Simoni:2011hydrologic}
S.~Simoni, S.~Padoan, D.~F. Nadeau, M.~Diebold, A.~Porporato, G.~Barrenetxea,
  F.~Ingelrest, M.~Vetterli, and M.~Parlange, ``Hydrologic response of an
  alpine watershed: Application of a meteorological wireless sensor network to
  understand streamflow generation,'' \emph{Water Resources Research}, vol.~47,
  no.~10, 2011.

\bibitem{win2018efficient}
M.~Z. Win, F.~Meyer, Z.~Liu, W.~Dai, S.~Bartoletti, and A.~Conti, ``Efficient
  multisensor localization for the internet of things: Exploring a new class of
  scalable localization algorithms,'' \emph{IEEE Signal Processing Magazine},
  vol.~35, no.~5, pp. 153--167, 2018.

\bibitem{simkovits2017navigation}
H.~Simkovits, A.~J. Weiss, and A.~Amar, ``Navigation by inertial device and
  signals of opportunity,'' \emph{Signal Processing}, vol. 131, pp. 280--287,
  2017.

\bibitem{Rockah:1987array-partI}
Y.~Rockah and P.~Schultheiss, ``Array shape calibration using sources in
  unknown locations--{Part I}: Far-field sources,'' \emph{IEEE Transactions on
  Acoustics, Speech, and Signal Processing}, vol.~35, no.~3, pp. 286--299,
  1987.

\bibitem{Rockah:1987array-partII}
------, ``Array shape calibration using sources in unknown locations--{Part
  II}: Near-field sources and estimator implementation,'' \emph{IEEE
  Transactions on Acoustics, Speech, and Signal Processing}, vol.~35, no.~6,
  pp. 724--735, 1987.

\bibitem{Weiss:1989array}
A.~J. Weiss and B.~Friedlander, ``Array shape calibration using sources in
  unknown locations-a maximum likelihood approach,'' \emph{IEEE Transactions on
  acoustics, speech, and signal processing}, vol.~37, no.~12, pp. 1958--1966,
  1989.

\bibitem{plinge2016acoustic}
A.~Plinge, F.~Jacob, R.~Haeb-Umbach, and G.~A. Fink, ``Acoustic microphone
  geometry calibration: An overview and experimental evaluation of
  state-of-the-art algorithms,'' \emph{IEEE Signal Processing Magazine},
  vol.~33, no.~4, pp. 14--29, 2016.

\bibitem{wang2015self}
L.~Wang, T.-K. Hon, J.~D. Reiss, and A.~Cavallaro, ``Self-localization of
  ad-hoc arrays using time difference of arrivals,'' \emph{IEEE Transactions on
  Signal Processing}, vol.~64, no.~4, pp. 1018--1033, 2015.

\bibitem{torgerson1952multidimensional}
W.~S. Torgerson, ``Multidimensional scaling: I. {T}heory and method,''
  \emph{Psychometrika}, vol.~17, no.~4, pp. 401--419, 1952.

\bibitem{kruskal1978multidimensional}
J.~B. Kruskal and M.~Wish, \emph{Multidimensional scaling}.\hskip 1em plus
  0.5em minus 0.4em\relax Sage, 1978, no.~11.

\bibitem{dokmanic2015euclidean}
I.~Dokmani\'{c}, R.~Parhizkar, J.~Ranieri, and M.~Vetterli, ``Euclidean
  distance matrices: essential theory, algorithms, and applications,''
  \emph{IEEE Signal Processing Magazine}, vol.~32, no.~6, pp. 12--30, 2015.

\bibitem{vanwynsberghe2016jasa}
\BIBentryALTinterwordspacing
C.~Vanwynsberghe, P.~Challande, J.~Marchal, R.~Marchiano, and F.~Ollivier, ``A
  robust and passive method for geometric calibration of large arrays,''
  \emph{The Journal of the Acoustical Society of America}, vol. 139, no.~3, pp.
  1252--1263, 2016. [Online]. Available:
  \url{https://doi.org/10.1121/1.4944566}
\BIBentrySTDinterwordspacing

\bibitem{Schonemann:1970wd}
P.~H. Sch{\"o}nemann, ``On metric multidimensional unfolding,''
  \emph{Psychometrika}, vol.~35, no.~3, pp. 349--366, 1970.

\bibitem{Crocco:2012eu}
M.~Crocco, A.~Bue, and V.~Murino, ``A bilinear approach to the position
  self-calibration of multiple sensors,'' \emph{IEEE Trans. Signal Process.},
  vol.~60, no.~2, pp. 660--673, 2012.

\bibitem{biswas2006semidefinite1}
P.~Biswas, T.-C. Lian, T.-C. Wang, and Y.~Ye, ``Semidefinite programming based
  algorithms for sensor network localization,'' \emph{ACM Transactions on
  Sensor Networks (TOSN)}, vol.~2, no.~2, pp. 188--220, 2006.

\bibitem{biswas2006semidefinite2}
P.~Biswas, T.-C. Liang, K.-C. Toh, Y.~Ye, and T.-C. Wang, ``Semidefinite
  programming approaches for sensor network localization with noisy distance
  measurements,'' \emph{IEEE transactions on automation science and
  engineering}, vol.~3, no.~4, pp. 360--371, 2006.

\bibitem{yang2009efficient}
K.~Yang, G.~Wang, and Z.-Q. Luo, ``Efficient convex relaxation methods for
  robust target localization by a sensor network using time differences of
  arrivals,'' \emph{IEEE transactions on signal processing}, vol.~57, no.~7,
  pp. 2775--2784, 2009.

\bibitem{vaghefi2013asynchronous}
R.~M. Vaghefi and R.~M. Buehrer, ``Asynchronous time-of-arrival-based source
  localization,'' in \emph{International Conference on Acoustics, Speech and
  Signal Processing (ICASSP)}.\hskip 1em plus 0.5em minus 0.4em\relax IEEE,
  2013, pp. 4086--4090.

\bibitem{wang2012semidefinite}
G.~Wang, Y.~Li, and N.~Ansari, ``A semidefinite relaxation method for source
  localization using {TDOA} and {FDOA} measurements,'' \emph{IEEE Transactions
  on Vehicular Technology}, vol.~62, no.~2, pp. 853--862, 2012.

\bibitem{alfakih1999}
A.~Y. Alfakih, A.~Khandani, and H.~Wolkowicz, ``Solving euclidean distance
  matrix completion problems via semidefinite programming,''
  \emph{Computational Optimization and Applications}, vol.~12, no.~1, p.
  13–30, 1999.

\bibitem{ding2010}
Y.~Ding, N.~Krislock, J.~Qian, and H.~Wolkowicz, ``Sensor network localization,
  euclidean distance matrix completions, and graph realization,''
  \emph{Optimization and Engineering}, vol.~11, no.~1, pp. 45--66, 2010.

\bibitem{ono2009blind}
N.~Ono, H.~Kohno, N.~Ito, and S.~Sagayama, ``Blind alignment of asynchronously
  recorded signals for distributed microphone array,'' in \emph{2009 IEEE
  Workshop on Applications of Signal Processing to Audio and Acoustics}.\hskip
  1em plus 0.5em minus 0.4em\relax IEEE, 2009, pp. 161--164.

\bibitem{zhang2016iwaenc}
J.~{Zhang}, R.~C. {Hendriks}, and R.~{Heusdens}, ``Structured total least
  squares based internal delay estimation for distributed microphone
  auto-localization,'' in \emph{2016 IEEE International Workshop on Acoustic
  Signal Enhancement (IWAENC)}, 2016, pp. 1--5.

\bibitem{Gaubitch:2013km}
N.~D. Gaubitch, W.~B. Kleijn, and R.~Heusdens, ``Auto-localization in ad-hoc
  microphone arrays,'' in \emph{IEEE ICASSP}.\hskip 1em plus 0.5em minus
  0.4em\relax Vancouver: IEEE, 2013, pp. 106--110.

\bibitem{thrun2006affine}
S.~Thrun, ``Affine structure from sound,'' in \emph{Advances in Neural
  Information Processing Systems}, 2006, pp. 1353--1360.

\bibitem{krekovic2018structure}
M.~Krekovi{\'c}, G.~Baechler, I.~Dokmani{\'c}, and M.~Vetterli, ``Structure
  from sound with incomplete data,'' in \emph{2018 IEEE International
  Conference on Acoustics, Speech and Signal Processing (ICASSP)}.\hskip 1em
  plus 0.5em minus 0.4em\relax IEEE, 2018, pp. 3539--3543.

\bibitem{peng2007beepbeep}
C.~Peng, G.~Shen, Y.~Zhang, Y.~Li, and K.~Tan, ``Beepbeep: a high accuracy
  acoustic ranging system using cots mobile devices,'' in \emph{Proceedings of
  the 5th international conference on Embedded networked sensor systems}.\hskip
  1em plus 0.5em minus 0.4em\relax ACM, 2007, pp. 1--14.

\bibitem{vanwynsberghe2016reseaux}
C.~Vanwynsberghe, ``R{\'e}seaux {\`a} grand nombre de microphones:
  applicabilit{\'e} et mise en {\oe}uvre,'' Ph.D. dissertation, Paris 6, 2016.

\bibitem{dokmanic2015relax}
I.~{Dokmani\'{c}}, J.~{Ranieri}, and M.~{Vetterli}, ``Relax and unfold:
  Microphone localization with euclidean distance matrices,'' in \emph{2015
  23rd European Signal Processing Conference (EUSIPCO)}, 2015, pp. 265--269.

\bibitem{pollefeys2008direct}
M.~Pollefeys and D.~Nister, ``Direct computation of sound and microphone
  locations from time-difference-of-arrival data,'' in \emph{2008 IEEE
  International Conference on Acoustics, Speech and Signal Processing}.\hskip
  1em plus 0.5em minus 0.4em\relax IEEE, 2008, pp. 2445--2448.

\bibitem{Heusdens:2014er}
R.~Heusdens and N.~Gaubitch, ``Time-delay estimation for {TOA}-based
  localization of multiple sensors,'' in \emph{IEEE ICASSP}.\hskip 1em plus
  0.5em minus 0.4em\relax IEEE, 2014, pp. 609--613.

\bibitem{jiang2013time}
F.~Jiang, Y.~Kuang \emph{et~al.}, ``Time delay estimation for {TDOA}
  self-calibration using truncated nuclear norm regularization,'' in
  \emph{International Conference on Acoustics, Speech and Signal Processing
  (ICASSP)}.\hskip 1em plus 0.5em minus 0.4em\relax IEEE, 2013, pp. 3885--3889.

\bibitem{kuang2013complete}
Y.~Kuang, S.~Burgess, A.~Torstensson, and K.~{\AA}str{\"o}m, ``A complete
  characterization and solution to the microphone position self-calibration
  problem,'' in \emph{International Conference on Acoustics, Speech and Signal
  Processing (ICASSP)}.\hskip 1em plus 0.5em minus 0.4em\relax IEEE, 2013, pp.
  3875--3879.

\bibitem{kuang2013stratified}
Y.~Kuang and K.~{\AA}str{\"o}m, ``Stratified sensor network self-calibration
  from {TDOA} measurements,'' in \emph{European Signal Processing Conference
  (EUSIPCO)}.\hskip 1em plus 0.5em minus 0.4em\relax IEEE, 2013, pp. 1--5.

\bibitem{zhayida2015toa}
S.~Zhayida, S.~Burgess, Y.~Kuang, and K.~{\AA}str{\"o}m, ``{TOA}-based
  self-calibration of dual-microphone array,'' \emph{Journal of Selected Topics
  in Signal Processing}, vol.~9, no.~5, pp. 791--801, 2015.

\bibitem{farmani2016tdoa}
M.~Farmani, R.~Heusdens, M.~S. Pedersen, and J.~Jensen, ``Tdoa-based
  self-calibration of dual-microphone arrays,'' in \emph{European Signal
  Processing Conference (EUSIPCO)}.\hskip 1em plus 0.5em minus 0.4em\relax
  IEEE, 2016, pp. 617--621.

\bibitem{burgess2015toa}
S.~Burgess, Y.~Kuang, and K.~{\AA}str{\"o}m, ``{TOA} sensor network
  self-calibration for receiver and transmitter spaces with difference in
  dimension,'' \emph{Signal Processing}, vol. 107, pp. 33--42, 2015.

\bibitem{batstone2019robust}
K.~Batstone, G.~Flood, T.~Beleyur, V.~Larsson, H.~R. Goerlitz, M.~Oskarsson,
  and K.~{\AA}str{\"o}m, ``Robust self-calibration of constant offset
  time-difference-of-arrival,'' in \emph{International Conference on Acoustics,
  Speech and Signal Processing (ICASSP)}.\hskip 1em plus 0.5em minus
  0.4em\relax IEEE, 2019, pp. 4410--4414.

\bibitem{Dokmanic2014tc}
I.~Dokmani{\'c}, L.~Daudet, and M.~Vetterli, ``How to localize ten microphones
  in one fingersnap,'' in \emph{EUSIPCO}, 2014.

\bibitem{Dokmanic2016gu}
------, ``From acoustic room reconstruction to {SLAM},'' in \emph{IEEE
  ICASSP}.\hskip 1em plus 0.5em minus 0.4em\relax IEEE, 2016, pp. 6345--6349.

\bibitem{sturm1999sedumi}
\BIBentryALTinterwordspacing
J.~F. Sturm, ``{Using SeDuMi 1.02, A Matlab toolbox for optimization over
  symmetric cones},'' \emph{Optimization Methods and Software}, vol.~11, no.
  1-4, pp. 625--653, 1999. [Online]. Available:
  \url{https://doi.org/10.1080/10556789908805766}
\BIBentrySTDinterwordspacing

\bibitem{boydalgebra}
S.~Boyd and L.~Vandenberghe, \emph{Introduction to Applied Linear Algebra:
  Vectors, Matrices, and Least Squares}.\hskip 1em plus 0.5em minus 0.4em\relax
  Cambridge University Press, 2018.

\bibitem{cvx}
M.~Grant and S.~Boyd, ``{CVX}: Matlab software for disciplined convex
  programming, version 2.1,'' \url{http://cvxr.com/cvx}, Mar. 2014.

\bibitem{gb08}
------, ``Graph implementations for nonsmooth convex programs,'' in
  \emph{Recent Advances in Learning and Control}, ser. Lecture Notes in Control
  and Information Sciences, V.~Blondel, S.~Boyd, and H.~Kimura, Eds.\hskip 1em
  plus 0.5em minus 0.4em\relax Springer-Verlag Limited, 2008, pp. 95--110,
  \url{http://stanford.edu/~boyd/graph_dcp.html}.

\bibitem{scheibler2018}
R.~Scheibler, E.~Bezzam, and I.~Dokmani\'{c}, ``Pyroomacoustics: A python
  package for audio room simulations and array processing algorithms,'' in
  \emph{IEEE International Conference on Acoustics, Speech and Signal
  Processing (ICASSP)}, 2018.

\bibitem{schoenemann1966}
\BIBentryALTinterwordspacing
P.~H. Sch{\"o}nemann, ``A generalized solution of the orthogonal procrustes
  problem,'' \emph{Psychometrika}, vol.~31, no.~1, pp. 1--10, 1966. [Online].
  Available: \url{https://doi.org/10.1007/BF02289451}
\BIBentrySTDinterwordspacing

\bibitem{speechcommands}
P.~Warden, ``Speech commands: A public dataset for single-word speech
  recognition,'' \emph{Dataset available from
  \url{http://download.tensorflow.org/data/speech_commands_v0.01.tar.gz}},
  2017.

\bibitem{yamaoka2019tdoa}
K.~{Yamaoka}, R.~{Scheibler}, N.~{Ono}, and Y.~{Wakabayashi}, ``Sub-sample time
  delay estimation via auxiliary-function-based iterative updates,'' in
  \emph{2019 IEEE Workshop on Applications of Signal Processing to Audio and
  Acoustics (WASPAA)}, 2019, pp. 130--134.

\end{thebibliography}

\end{document}